\documentclass[11pt]{article}
\usepackage{epsfig} 
\usepackage{amssymb}
\newcommand{\sect}[1]{ \section{#1} \setcounter{equation}{0} }

\newcommand{\kslash}{k \! \! \! /}

\newcommand{\Deltaslash}{\Delta \! \! \! \! / \,}

\newcommand{\half}{\mbox{\small{$\frac{1}{2}$}}}
\newcommand{\third}{\mbox{\small{$\frac{1}{3}$}}} 
 
\newcommand{\threehalves}{\mbox{\small{$\frac{3}{2}$}}} 
\newcommand{\pitwo}{\mbox{\small{$\frac{\pi}{2}$}}} 
\newcommand{\pisix}{\mbox{\small{$\frac{\pi}{6}$}}} 
\newcommand{\MSbar}{\overline{\mbox{MS}}}

\newcommand{\Nf}{N_{\!f}}

\newcommand{\Cc}{\mathbb{C}}
\newcommand{\Gg}{\mathbb{G}}

\begin{document}
\title{Three loop renormalization of $3$-quark operators in QCD}
\author{J.A. Gracey, \\ Theoretical Physics Division, \\ 
Department of Mathematical Sciences, \\ University of Liverpool, \\ P.O. Box 
147, \\ Liverpool, \\ L69 3BX, \\ United Kingdom.} 
\date{} 
\maketitle 

\vspace{5cm} 
\noindent 
{\bf Abstract.} We compute the three loop $\MSbar$ anomalous dimension for the
$3$-quark operator corresponding to the proton. This requires the treatment of
$\gamma^5$ within dimensional regularization as well as evanescent operators 
generated through the renormalization. We extend the Larin scheme for
$\gamma^5$ to a mixing matrix of finite renormalization constants chosen so 
that chiral symmetry is manifest in four dimensions. We also provide the finite
part of the Green's function at two loops where the operator is inserted at 
zero momentum in a quark $3$-point function in an arbitrary linear covariant 
gauge in order to assist with the lattice measurement of the same quantity. The
renormalization of the generalized $3$-quark operators in the scheme devised by
Kr\"{a}nkl and Manashov is extended to three loops and the anomalous dimensions
for the $(\half,0)$, $(\threehalves,0)$ and $(1,\half)$ spin operators with 
various chiralities are also given. 

\vspace{-19.2cm}
\hspace{13.5cm}
{\bf LTH 957}

\newpage

\sect{Introduction.}

In quantum field theory baryons are represented by $3$-quark operators where
the operators are chosen so as to have the same discrete and continuous
symmetries as the observed states in nature guided by the quark model. For 
instance, protons are built from two up and one down quark fields with an 
overall spin of spin $\half$. With suitable choices of handedness for these 
three fields one can construct a $3$-quark operator with the correct $J^{PC}$ 
values for the proton. While in nature these quarks are in general massive and 
confined, for some theoretical studies of such hadronic states one can consider
them to be built from chiral or massless fields. This is an appropriate 
approximation in the high energy limit where the masses are small when compared
to the momentum scale. Thus the leading twist operators will dominate any high 
energy analysis, \cite{1,2,3}. At low energies the internal structure of the 
baryons and in particular protons, can be probed by measuring the structure 
functions and their moments. For the latter the field theoretic quantities of 
relevance are $3$-quark operators involving covariant derivatives, 
\cite{1,2,3,4,5,6}. Knowledge of the proton structure can assist with 
understanding the strong interaction in the infrared regime and hence how 
quarks condense or hadronize to form the nucleon states we see in nature. In 
quantum field theory the use of the Quantum Chromodynamics (QCD) Lagrangian 
provides us with a forum to study these different aspects of hadrons. At high 
energy one can apply perturbation theory and calculate order by order in the 
strong coupling constant which is assumed to be small in this region. However, 
while this provides one with the renormalization group evolution of the 
operators with scale, it is the non-perturbatively measured operator matrix 
element which gives the information relating to the baryon structure. Such 
matrix elements cannot be determined perturbatively for the purpose of 
extracting physical results. Instead they are measured non-perturbatively using
lattice gauge theory where the spacetime is discretized. Although this involves
the use of large computers to handle the huge numerical calculations, there are 
various technical issues underlying the process. One of these resides in making
accurate measurements through credibly small error bars. Moreover, one needs to
have contact with the continuum structure of the same quantities once the 
lattice regularization is lifted. Indeed it should be the case that when low 
energy estimates from the lattice are extrapolated to the high energy continuum
evaluation there should be reasonable agreement. 

To this end over a period of years there has been ongoing comparison of lattice
computations with high energy perturbative expressions. The latter are 
determined in the chiral limit, usually in the $\MSbar$ scheme, to as high a 
loop order as is calculationally possible. Indeed with the current loop 
calculation technology this invariably means {\em three} loops through the use 
of the {\sc Mincer} package, \cite{7,8}. As highlights of this bridge we 
mention quark current renormalization for zero momentum operator insertions,
\cite{9,10}, and more recently at non-zero momentum insertion, 
\cite{11,12,13,14,15}. The latter are known at two loops at the symmetric 
subtraction point for the $\MSbar$ scheme. However, in both cases results have 
been determined in the regularization invariant (RI), \cite{16,17}, class of
schemes which includes RI${}^\prime$ and RI${}^\prime$/SMOM, \cite{11}, where 
SMOM indicates momentum subtraction at the symmetric point. These are schemes 
which are devised for lattice regularization in order to minimize the use of 
derivatives, which is computationally intense, in extracting renormalization 
constants as well as amplitudes. Both sets of schemes have continuum analogues,
\cite{9,10}. Building on this project it is natural now to turn from 
quark-antiquark operators representing mesonic physics to $3$-quark operators 
for baryon problems. Therefore, it is the purpose of this article to compute 
the Green's function for the zero momentum insertion of the $3$-quark operator 
representing a proton in the chiral limit to two loops in the $\MSbar$ scheme. 
In addition to assist with running to high energy we will compute the operator 
anomalous dimension to {\em three} loops in the same scheme. Both results will 
therefore be important for matching lattice results of the same Green's 
function in the high energy limit. For the operators which we consider here, we
restrict ourselves to the $\MSbar$ scheme and will not introduce any variant of
the regularization invariant class of renormalization schemes. Though from the 
results compiled here it is possible to define such schemes for the $3$-quark 
operator renormalization and then convert to the $\MSbar$ scheme. However, one
reason for not choosing to explicitly include, say, RI${}^\prime$ results rests
in the fact that as noted in \cite{14} the definition of such schemes is not 
unique for Lorentz structure beyond the simplest quark mass operator. Indeed in
the tensor current case a scheme alternative, \cite{14}, to the original one of 
\cite{11,13} appeared to converge quicker at two loops. Only a three loop 
calculation would give more insight into this. 

In more detail we will evaluate the two loop Green's function for an arbitrary
linear covariant gauge at a point where the squared momenta of the three
external quark legs are all equal to the same non-zero value. Hence, there is
no zero momentum external quark leg. This is appropriate since a zero momentum
quark would be difficult to incorporate on the lattice. Several lattice studies
of $3$-quark operators and their low moments for similar Green's functions 
deserve mention. In \cite{18} an initial one loop lattice analysis was 
performed motivated by proton decay in an $SU(5)$ grand unified theory. More 
recently the QCDSF collaboration extended aspects of that analysis to a full
examination of the $3$-quark {\em lattice} operators including moments, 
\cite{19}. Though in both cases precise measurements require continuum 
perturbation theory for matching. While the lattice ultimately will only 
require results in the Landau gauge, we choose an arbitrary gauge for internal 
checking purposes. For instance, in extracting the anomalous dimension of the 
inserted $3$-quark operator, the result has to be independent of the gauge 
parameter in the $\MSbar$ scheme. This will represent a useful check here. 
Previously the two loop  $\MSbar$ anomalous dimensions were computed in the 
Feynman gauge in \cite{20}. The initial one loop analysis was carried out in 
\cite{1,2,3}. In finding total agreement with the expressions of 
\cite{1,2,3,20} at two loops we will have another check on our results. At 
three loops to reduce the computation time for the large number of Feynman 
graphs to be evaluated, we will restrict the calculation at that order to the 
Feynman gauge. Though the way that part proceeds there will be an internal 
check over and above that provided by the renormalization group equation. It is
also worth noting that the anomalous dimensions of $3$-quark operators have 
been determined to two loops in heavy quark effective theory, \cite{21,22}. In 
that analysis one of the three quarks in the operator is regarded as having a 
mass significantly larger than the other two. Equally $3$-quark operators have 
been used to estimate baryon masses using the operator product expansion, 
\cite{23,24,25,26,27,28,29}.

While this summary covers our aims it would be remiss at this stage not to
mention several technical problems which have to be addressed. The $3$-quark
operators share a similar feature to four-fermi operators. Not only can they 
mix under renormalization but within dimensional regularization in
$d$~$=$~$4$~$-$~$2\epsilon$ spacetime dimensions, which we use throughout and 
where $\epsilon$ is the regularizing parameter, the operators will mix into
evanescent operators. These exist in the analytically continued spacetime but
are non-existent in strictly four dimensions. However, their presence within
the renormalization has to be taken into account. We will use the projection
technique of \cite{30}. Coupled to this is the underlying $\gamma^5$ issue
due to the fermion handedness. For automatic symbolic manipulation calculations
a method was developed in \cite{31} to accommodate $\gamma^5$ with operator
renormalization. We adopt and adapt that technique for the $3$-quark operators
relating to the proton. At two loops the $\gamma^5$ problem does not arise in
$\MSbar$ for the operator anomalous dimension due to the nature of the mixing
matrix. At three loops the problem will be evident and needs to be treated. 

Finally, we will extend a more recent two loop renormalization of a $3$-quark
operator which was introduced in \cite{32}. In \cite{32} rather than use an 
initial operator with the correct quantum numbers the most general $3$-quark 
operator, devoid of any $\gamma$-matrix structure, was renormalized using 
dimensional regularization and an $\MSbar$ scheme subtraction. Clearly the seed
operator will mix into a set of $3$-quark operators with $\gamma$-matrices. 
These arise from the internal quark propagators and quark-gluon vertices of the
Feynman diagrams at each order. The claim in \cite{32} is that from this mixing
one can overcome not only $\gamma^5$ issues but also handle the evanescent 
operator problem in a systematic way. This is motivated by earlier work of
\cite{33}. Hence we will first check the two loop general anomalous dimension 
of \cite{32} in an arbitrary linear covariant gauge and then extend the result 
of \cite{32} to three loops. The former will also play the role of a subsidiary
check on our proton operator $\MSbar$ result. Though it is worth stressing that
the scheme dependent parts of the two loop results of \cite{32} are not the 
same as those of \cite{20} which is regarded as being the $\MSbar$ scheme. As 
noted in \cite{32} they have provided a conversion factor which appears to 
derive from $O(\epsilon)$ corrections similar to contributions from evanescent 
terms. However, one advantage of the general operator of \cite{32} is that the 
anomalous dimension of $3$-quark operators of the spin-$(j,\bar{j})$ Lorentz 
type can be easily deduced from the general anomalous dimension. Hence we will 
provide the three loop value for, say, the Ioffe current \cite{23} among other 
quantities. 

The article is organized as follows. Section $2$ is devoted to the background 
for the renormalization of the operators we will examine at three loops 
including the definition of the operator basis in $d$-dimensions. We provide 
technical details of the calculation of the Green's functions of interest in 
both momentum configurations in section $3$. The results of the renormalization
are recorded in section $4$ including the finite renormalization required to 
treat $\gamma^5$. The explicit form of the two loop amplitude relevant for 
lattice matching is presented in section $5$. Section $6$ is devoted to the 
extension of \cite{32} to three loops where we record the anomalous dimensions 
of various spin operators at three loops in the scheme used in \cite{32}.
Finally, we conclude with remarks in section $7$.

\sect{Background.}

We begin by first discussing the actual operators we will renormalize. For
further background we note that we have based this on the analysis of 
\cite{20}. Clearly the proton involves up and down quarks and the relevant 
operators in the chiral limit with the correct parity are  
\begin{eqnarray}
{\cal O}^{udu}_1 &=& \epsilon^{IJK} \gamma^5 u^I \left( \left( u^J \right)^T C 
d^K \right) \nonumber \\
{\cal O}^{udu}_2 &=& \epsilon^{IJK} u^I \left( \left( u^J \right)^T C \gamma^5 
d^K \right) 
\label{origop}
\end{eqnarray} 
which carry one free spinor index. Here $C$ is the charge conjugation matrix 
which satisfies $C C$~$=$~$-$~$1$ with 
\begin{equation}
C \left( \gamma^\mu \right)^T C ~=~ \gamma^\mu ~.
\label{Cgam}
\end{equation}
The indices $I$, $J$ and $K$ are $SU(3)$ colour indices and throughout we work
in this specific Lie group. Therefore, within our calcluations products of the 
group generators are simplified automatically with  
\begin{equation}
T^a_{IJ} T^a_{KL} ~=~ \frac{1}{2} \left[ \delta_{IL} \delta_{KJ}
- \frac{1}{3} \delta_{IJ} \delta_{KL} \right] 
\end{equation}
where $1$~$\leq$~$a$~$\leq$~$8$.
This means that when, for example, we make use of the quark wave function and 
gauge parameter anomalous dimensions as well as the QCD $\beta$-function then 
the usual group Casimirs, $C_F$, $C_A$ and $T_F$, of those expressions are also 
evaluated at their $SU(3)$ values. Indeed we stress that the $3$-quark 
operators we consider here have no physical meaning for colour groups other 
than $SU(3)$ since only for that group are they gauge invariant.

In (\ref{origop}) we have two operators of the same dimension and discrete 
symmetries. These will mix under renormalization. However, various linear 
combinations will produce the correct number of left and right handed quarks to
produce the operator which actually corresponds to the proton itself. As was 
discussed in \cite{20} for practical computational purposes it is more 
appropriate to use a related set of operators to perform our two and three loop
calculations. These operators are given by 
\begin{eqnarray}
{\cal O}_1 &=& \epsilon^{IJK} \psi^I \left( \left( \psi^J \right)^T C 
\psi^K \right) \nonumber \\
{\cal O}_2 &=& \epsilon^{IJK} \gamma^5 \psi^I \left( \left( \psi^J \right)^T C 
\gamma^5 \psi^K \right) ~, 
\label{renop}
\end{eqnarray} 
where we have omitted the flavour indices, and are related to the operators of 
(\ref{origop}) by  
\begin{equation}
{\cal O}_i ~=~ \gamma^5 {\cal O}^{udu}_i ~.
\end{equation}
From now on our focus will be on ${\cal O}_i$. Though we note at this point
that whichever set of operators one uses they are gauge invariant for $SU(3)$ 
colour and so in the $\MSbar$ scheme, which we will use, the anomalous
dimensions will be independent of the gauge parameter. This will be checked
explicitly to two loops. The original computations of \cite{20} were performed
in the Feynman gauge rather than the arbitrary linear covariant gauge we use
here to two loops.

For instance, as we have two operators of the same dimension and symmetries
their mixing under renormalization is handled by a mixing matrix of 
renormalization constants and thence a matrix of anomalous dimensions. For the
former we therefore have  
\begin{equation}
{\cal O}_{i\,\mbox{\footnotesize o}} ~=~ Z_{ij} {\cal O}_j
\end{equation}
where the subscript ${}_{\mbox{\footnotesize o}}$ denotes the bare operator.
In our conventions the matrix of anomalous dimensions is defined by
\begin{equation}
\gamma_{ij} (a) ~=~ -~ \mu \frac{d ~}{d \mu} \ln Z_{ij} 
\label{matdef}
\end{equation}
where 
\begin{equation}
\mu \frac{d~}{d\mu} ~=~ \beta(a) \frac{\partial ~}{\partial a} ~+~
\alpha \gamma_\alpha(a,\alpha) \frac{\partial ~}{\partial \alpha} 
\end{equation}
and $\mu$ is the renormalization scale introduced to ensure that the coupling
constant, $g$, is dimensionless in $d$-dimensions. Our gauge parameter is 
$\alpha$ with $\alpha$~$=$~$0$ corresponding to the Landau gauge. We choose to 
work with the coupling constant $a$ which is related to the gauge coupling 
constant and the strong force coupling constant, $\alpha_s$, by
\begin{equation}
a ~=~ \frac{g^2}{16\pi^2} ~~~~~,~~~~~
a ~=~ \frac{\alpha_s}{4\pi} ~.
\end{equation} 
For mass dependent renormalization schemes $\gamma_{ij}(a)$ can depend on
$\alpha$ but we have omitted any $\alpha$ dependence on the left side of
(\ref{matdef}) since we focus on $\MSbar$. For practical purposes in the
extraction of the anomalous dimensions we note that (\ref{matdef}) implies
\begin{equation}
\gamma_{ik}(a) Z_{kj} ~=~ Z_{ik} \gamma_{kj}(a) ~=~ -~ \mu \frac{d ~}{d \mu} 
Z_{ij} 
\end{equation} 
and we have checked that the same mixing matrix emerges irrespective of which
way the matrices are multiplied. 

It should be stressed that this discussion about the structure of the mixing 
matrix, where $i$ and $j$ run over $1$ and $2$, is in essence the situation in
four spacetime dimensions. However, as will be apparent later since we will be
using dimensional regularization the mixing matrix is not finite dimensional in
$d$-dimensions. It will have to be extended to an infinite dimensional case at
all orders in perturbation theory. This is because in $d$-dimensions operators 
will be generated through the renormalization which have no strictly four 
dimensional equivalent. Put another way their four dimensional equivalent is 
the zero operator and so such operators in $d$-dimensions are referred to as 
being evanescent. Their presence in this analysis is not solely because of the 
presence of $\gamma^5$. These evanescent operators would arise if there was no 
mixing and one was only considering ${\cal O}_1$. However, both the evanescent 
and $\gamma^5$ issues will have to be dealt with at the appropriate point. 
While we have noted that $Z_{ij}$ will become infinite dimensional in 
$d$-dimensions, it will do so in a controlled way in that at each order in 
perturbation theory the matrix will be extended by the appearance of a new 
operator. In other words at each order in perturbation theory the matrix 
enlarges but remains finite dimensional. 

To appreciate these remarks we need to focus on the treatment of 
$\gamma$-matrices. In $d$-dimensions one has to extend the basis of 
$\gamma$-matrices to an infinite set of matrices denoted by 
$\Gamma_{(n)}^{\mu_1\ldots\mu_n}$, \cite{30,34,35}, which are totally 
antisymmetric in the Lorentz indices and defined by 
\begin{equation}
\Gamma_{(n)}^{\mu_1\ldots\mu_n} ~=~ \gamma^{[\mu_1} \ldots \gamma^{\mu_n]}
\end{equation}
where a factor of $1/n!$ is understood and $n$ is an integer, $n$~$\geq$~$0$. 
These generalized matrices span spinor space in $d$-dimensions and the 
underlying algebra necessary for loop calculations has been developed in 
various articles, \cite{30,34,35}. For instance, the trace operation is 
isotropic with respect the basis since,
\cite{36,37},
\begin{equation}
\mbox{tr} \left( \Gamma_{(m)}^{\mu_1 \ldots \mu_m}
\Gamma_{(n)}^{\nu_1 \ldots \nu_n} \right) ~ \propto ~ \delta_{mn}
I^{\mu_1 \ldots \mu_m \nu_1 \ldots \nu_n} ~.
\end{equation}
Moreover, it is possible to write products of the original $\gamma$-matrices as
a finite sum over $\Gamma_{(n)}^{\mu_1\ldots\mu_n}$. This can be achieved
recursively by applying the relations, \cite{35,36,37}, 
\begin{eqnarray}
\Gamma^{\mu_1 \ldots \mu_n}_{(n)} \gamma^\nu &=&
\Gamma^{\mu_1 \ldots \mu_n \nu}_{(n+1)} ~+~ \sum_{r=1}^n (-1)^{n-r} \,
\eta^{\mu_r \nu} \, \Gamma^{\mu_1 \ldots \mu_{r-1} \mu_{r+1} \ldots
\mu_n}_{(n-1)} \\
\gamma^\nu \Gamma^{\mu_1 \ldots \mu_n}_{(n)} &=&
\Gamma^{\nu \mu_1 \ldots \mu_n}_{(n+1)} ~+~ \sum_{r=1}^n (-1)^{r-1} \,
\eta^{\mu_r \nu} \, \Gamma^{\mu_1 \ldots \mu_{r-1} \mu_{r+1} \ldots
\mu_n}_{(n-1)} 
\label{gamnalg}
\end{eqnarray}
where $\eta_{\mu\nu}$ is the spacetime metric tensor. For reference when one 
restricts the $\Gamma_{(n)}$-matrices to four dimensions we have 
\begin{eqnarray} 
\left. \Gamma_{(2)}^{\mu\nu} \right|_{d=4} &=& \sigma^{\mu\nu} ~~~,~~~
\left. \Gamma_{(4)}^{\mu\nu\sigma\rho} \right|_{d=4} ~=~ 
\epsilon^{\mu\nu\sigma\rho} \gamma^5 \nonumber \\
\left. \Gamma_{(n)}^{\mu_1\ldots\mu_n} \right|_{d=4} &=& 0 ~~ 
\mbox{for $n$~$\geq$~$5$} 
\label{gammap}
\end{eqnarray}
where $\epsilon^{\mu\nu\sigma\rho}$ is the four dimensional totally 
antisymmetric pseudotensor. As a note we mention that the $\gamma^5$ matrix 
which exists in strictly four dimensions and defines chirality has absolutely 
no connection whatsoever with $\Gamma_{(5)}^{\mu_1\ldots\mu_5}$ in 
$d$-dimensions. The former object exists only in four dimensions and Larin's
procedure, \cite{31}, which we use to handle $\gamma^5$ within dimensional 
regularization will be detailed later for the present computation. As a 
notational comment we will use $\gamma^\mu$ and $\Gamma_{(1)}^\mu$ synonymously
in $d$-dimensions since in (\ref{gamnalg}) as the former is less clumsy but 
regard $\sigma^{\mu\nu}$ as the purely four dimensional object. 

There are two sources of these generalized $\gamma$-matrices within the 
computations. The first is the simplest and alluded to already and that is that
the product of $\gamma$-matrices which remain in the calculation of either
Green's function can be written in the general basis. This leads to the second
source which is the generation of operators into which the seed operator,
(\ref{renop}), of the Green's function mixes under renormalization. In 
$d$-dimensions given that we have to use the generalized basis we then have to 
use the generalization of the four dimensional operators to the same basis. 
Therefore we define the new $d$-dimensional operators
\begin{eqnarray}
{\cal O}_{(n)} &=& \epsilon^{IJK} 
\left( \Gamma_{(n)}^{\mu_1\ldots\mu_n} \psi^I \right) 
\left( \left( \psi^J \right)^T C \Gamma_{(n)\,\mu_1\ldots\mu_n} \psi^K 
\right) ~~~,~~~ \mbox{for $n$~$\neq$~$4$} \nonumber \\
{\cal O}_{(4)} &=& \frac{1}{24} \epsilon^{IJK} 
\left( \Gamma_{(4)}^{\mu_1\mu_2\mu_3\mu_4} \psi^I \right)
\left( \left( \psi^J \right)^T C \Gamma_{(4)\,\mu_1\mu_2\mu_3\mu_4} \psi^K 
\right) ~~~,~~~ 
\mbox{for $n$~$=$~$4$}
\label{opbas}
\end{eqnarray}
where the factor for ${\cal O}_{(4)}$ is to ensure that there is a direct
mapping in the four dimensional limit to the original operators. This follows
from (\ref{gammap}) and the fact that in $\epsilon_{\mu\nu\sigma\rho}
\epsilon^{\mu\nu\sigma\rho}$~$=$~$24$. Hence,
\begin{equation}
{\cal O}_{(4)} ~=~ {\cal O}_2 ~+~ O(\epsilon)
\end{equation}
in the limit to four dimensions. As a mnemonic we note that
\begin{equation}
{\cal O}_n ~=~ {\cal O}_{(4n-4)}
\end{equation}
for $n$~$\geq$~$1$ and 
\begin{eqnarray}
\left. {\cal O}_1 \right|_{d=4} &=& \epsilon^{IJK} \psi^I \left( \left( \psi^J 
\right)^T C \psi^K \right) \nonumber \\
\left. {\cal O}_2 \right|_{d=4} &=& \epsilon^{IJK} \left( \gamma^5 \psi^I
\right) \left( \left( \psi^J \right)^T C \gamma^5 \psi^K \right) \nonumber \\
\left. {\cal O}_n \right|_{d=4} &=& 0 ~~~ \mbox{for $n$~$\geq$~$3$} ~.
\end{eqnarray}
Thus at the outset we are forced to consider a more general renormalization
from which the anomalous dimensions of the original operators will emerge as a
corollary. So within all our computations we will write the Green's functions
in terms of $\Gamma_{(n)}$-matrices. For the finite parts, which is ultimately
required for lattice matching, the Lorentz indices on these matrices can be
contracted with the external momenta. Though such objects will not be 
associated with a divergence in $\epsilon$ since this would violate the
renormalizability of the operators. Before restricting the $d$-dimensional 
Green's function there are more channels. This is one of the reasons why we 
have not taken the projection approach which was the main tool for mesonic 
operators, \cite{14,15}. For instance, one has to have knowledge of the full 
basis to say three loops for (\ref{setup2}) and then construct the projection 
tensor to isolate all the possible spinor channels. Within a symbolic 
manipulation approach this would significantly reduce run times due, in part, 
to having to internally manipulate products of $\Gamma_{(n)}$-matrices. Instead
we have constructed the relations between the products of 
$\Gamma_{(m)}^{\mu_1\ldots\mu_m}$ and $\Gamma_{(n)}^{\nu_1\ldots\nu_n}$ for 
various values of $m$ and $n$ which occur. These have been encoded within a 
module in the symbolic manipulation language, {\sc Form}, \cite{38} which we 
use throughout. Such a product can be written in terms of 
$\Gamma_{(p)}^{\sigma_1\ldots\sigma_p}$ where $|m-n|$~$\leq$~$p$~$\leq$~$(m+n)$
and $\sigma_i$~$\in$~$\{\mu_1,\ldots,\mu_m,\nu_1,\ldots,\nu_n\}$. The indices
not used from this set in $\Gamma_{(p)}^{\sigma_1\ldots\sigma_p}$ appear in the
$\eta^{\mu\nu}$ tensors which are required to keep the total number of free
indices of each term as $(m+n)$.

Given the appearance now of the full basis of operators ${\cal O}_{(n)}$ in
$d$-dimensions we will determine the associated renormalization constant matrix
$Z_{ij}$. This matrix will increase in size at each loop order but from it we 
will determine what we will refer to as the naive anomalous dimension matrix,
$\tilde{\gamma}_{ij}(a)$. It would ordinarily correspond to the correct four
dimensional anomalous dimensions but in using these generalized operators we
have ignored the problem of $\gamma^5$ as well as the effect of the 
evanescent operators. The latter affect the structure of the four dimensional
anomalous dimensions even though the evanescent operators are non-existent in
four spacetime dimensions. To account for this we note the formalism 
developed in \cite{30} which appends to the naive anomalous dimensions extra
contributions which derive from the evanescent parts. We make minimal comment 
on the technique here since it transpires that the effect they have on the 
proton operator renormalization will not occur until {\em four} loops. Though
if there were contributions these together with the naive anomalous dimensions
would contribute to the correct four dimensional result. We say contribute as
one has also to deal with the absence of chiral symmetry in $d$-dimensions
which has been ignored with the choice of ${\cal O}_{(4)}$. Including the
evanescent effects with the naive anomalous dimensions for the present case
would not produce a result consistent with chiral symmetry in strictly four
dimensions. The procedure we have chosen to do this is based on Larin's
method, \cite{31}, which was developed for flavour non-singlet and singlet
quark currents as well as the chiral anomaly. We will discuss the technical
aspects of the calculation for our case later but in essence one needs to 
append a {\em finite} renormalization constant to the naive renormalization
constant. It is chosen in such a way that in strictly four dimensions the
anti-commutativity of $\gamma^5$ with $\gamma^\mu$ is restored. This aspect is
treated {\em after} the contributions from the evanescent operators have been
included. The criterion for defining the condition will be similar for our 
operators ${\cal O}_i$, $i$~$=$~$1$ and $2$, and like \cite{31} will be derived
from knowledge of the finite part of the Green's functions with the operators
inserted. Though the finite renormalization can also be derived from the 
difference in anomalous dimensions. However, in either case unlike \cite{31} it
will be a matrix of finite renormalization constants leading to an additional 
matrix of anomalous dimensions, $\gamma_{5,ij}(a)$. Thus the correct four 
dimensional anomalous dimension matrix for (\ref{origop}) or (\ref{renop}) will
formally be 
\begin{equation}
\gamma_{ij}(a) ~=~ \tilde{\gamma}_{ij}(a) ~+~ \gamma_{5,ij}(a) 
\end{equation}
where we exclude any evanescent part for this case. Here 
\begin{equation}
\gamma_{5,ij}(a) ~=~ -~ \mu \frac{d~}{d\mu} Z^5_{ij}
\label{finmatdef}
\end{equation}
and $Z^5_{ij}$ is the finite renormalization constant matrix. If the original 
seed operators had included additional $\gamma$-matrices, with or without free
Lorentz indices such as those of \cite{24,39}, then the evanescent operator
contribution could occur at the three loop order we are interested in here. 
In writing $\gamma_{5,ij}(a)$ we are assuming that the finite renormalization
is independent of the gauge parameter. In the cases examined by Larin in
\cite{31} and in a more recent analysis of diquark operators, \cite{40}, the 
finite renormalization did not depend on $\alpha$ which is what we found here.
Indeed it is worth noting that as this additional piece corresponds to a finite
renormalization, for the {\em three} loop mixing matrix in four dimensions one 
only requires $Z^5_{ij}$ to {\em two} loops because of the presence of 
$\beta(a)$ in (\ref{finmatdef}).

We close this section by summarizing the two calculations we carry out. First,
in each case a $3$-quark operator is inserted at zero momentum into a quark 
$3$-point function. For illustration if for the moment we denote this generic 
operator by ${\cal O}$ then we will calculate
\begin{equation} 
\left. \frac{}{} \left\langle \psi_\alpha(p) \psi_\beta(q) \psi_\gamma(r)
{\cal O}_\delta(0) \right\rangle \right|_{p^2=q^2=r^2=-\mu^2}
\label{setup1}
\end{equation} 
to two loops to the finite part for arbitrary gauge parameter $\alpha$ in the
$\MSbar$ scheme. This is to assist with lattice matching to the same quantity
in the Landau gauge. For this case there are $3$ one loop and $40$ two loop
Feynman graphs to compute and we will call this the symmetric setup. In 
(\ref{setup1}) momentum conservation implies
\begin{equation}
r ~=~ -~ p ~-~ q
\end{equation}
and we use a symmetric subtraction point for the external legs
\begin{equation}
p^2 ~=~ q^2 ~=~ r^2 ~=~ -~ \mu^2
\label{symmpt}
\end{equation}
which implies
\begin{equation}
pq ~=~ \frac{1}{2} \mu^2 ~.
\label{symmptpq}
\end{equation}
We have chosen the subtraction point to be $(-\mu^2)$ so as to omit logarithms
in the finite part of the Green's function. In order to extract the three loop 
mixing matrix of anomalous dimensions we have to consider a different momentum 
configuration so that the {\sc Mincer} algorithm, \cite{7}, can be applied. We 
will refer to this as the {\sc Mincer} setup. In this case the Green's function
is 
\begin{equation} 
\left. \frac{}{} \left\langle \psi_\alpha(p) \psi_\beta(-p) \psi_\gamma(0)
{\cal O}_\delta(0) \right\rangle \right|_{p^2=-\mu^2} 
\label{setup2}
\end{equation} 
and there are $784$ three loop diagrams to determine in addition to those noted
earlier at lower order. In both situations, (\ref{setup1}) and (\ref{setup2}),
the Feynman diagrams are generated automatically using the {\sc Qgraf} package,
\cite{41}, before being converted into {\sc Form}, \cite{38}, input notation. 
The latter is the symbolic manipulation language we use and the full automatic 
computation is written in terms of it. Though the compilation of the 
expressions for each Feynman diagram for both setups were run with the threaded
version, {\sc Tform}, \cite{42}. The nullification of an external quark leg 
momentum in effect produces a $2$-point function which {\sc Mincer} requires. 

\sect{Computational technicalities.}

We devote this section to the various technical aspects of the computation. In 
order to evaluate the Green's function for the full momentum or symmetric
configuration we use the Laporta algorithm, \cite{43}. Each of the one and two 
loop Feynman graphs we have to evaluate involve strings of $\gamma$-matrices 
which have external and internal momenta embedded within them. We have
proceeded by writing all integrals in terms of scalar integrals. By this we 
mean integrals where there are at most scalar products of the momenta and the 
strings of $\gamma$-matrices have contractions with only external momenta. In
other words these only play a passive role in the subsequent evaluation and are
written in terms of the generalized $\Gamma_{(n)}$-matrices. To achieve this we
make a projection of the tensor integrals onto a basis of tensors built from
$\eta_{\mu\nu}$, $p_\mu$ and $q_\mu$. For the tensor reduction we use we have 
at most a rank $5$ tensor built from the two internal loop momenta at two 
loops. This is due to the fact that we are computing for an arbitrary linear 
covariant gauge. A Feynman gauge computation would be more compact but would 
have limited applicability for the lattice. Once the scalarized integrals have 
been determined these are rewritten purely in terms of the propagators of the 
graph and any additional propagators which are not part of the topology, 
\cite{43}. These latter propagators are required for irreducible scalar 
products but are chosen in such a way as to cover all possible scalar products 
of the internal momenta with themselves and the external momenta. In this form 
one applies the Laporta reduction. This is an algorithm which systematically 
constructs all the integration by parts relations and optionally the Lorentz 
identities between all the integrals which are needed. From this tower of 
relations it is possible to algebraically relate all the scalar integrals to a 
base set of master integrals. These are evaluated by explicit integration and 
thus the evaluation of the Feynman graph is complete.

In describing the general procedure we note that for practical purposes one has
to use computers to implement the Laporta algorithm. We have used {\sc Reduze},
\cite{44}, which uses the {\sc GiNaC} computer algebra system, \cite{45}, and 
is written in C$++$. For the one and two loop graphs we need to evaluate it 
transpires that there are three basic topologies. There is one at one loop and 
two at two loops. For the latter one is the ladder graph and the other is the 
non-planar two loop $3$-point graph. These and their extension to include one 
other propagator are sufficient to cover all possible irreducible tensor 
integrals. Thus using the {\sc Reduze} package we have created a database of 
relations covering all possible levels of scalar integrals which can arise. The
ones which are explicitly required are extracted and converted into a 
{\sc Form} module which is called at the appropriate point of the automatic 
computation. At the end the explicit expressions for the master integrals are 
substituted. The ones we use are summarized in \cite{13} but were evaluated in 
various articles, \cite{46,47,48,49}, using a variety of techniques.

For the {\sc Mincer} situation, (\ref{setup2}), the momentum configuration with
one nullified external quark leg could potentially introduce spurious infrared 
infinities which would need infrared rearrangement. However, this does not 
arise. This would be the case if the Feynman integral produced propagators such
as $1/(k^2)^2$, where $k$ is an internal loop momentum, which are infrared 
singular. These are absent because the quark propagator retains $\kslash$ in 
the numerator or the triple gluon or ghost vertices carry a momentum to lift 
the potential infrared singularity. By contrast to (\ref{setup1}) we take a 
more general operator in order to compute the anomalous dimension. In 
particular we seed the Green's function with 
\begin{equation}
{\cal O} ~=~ \epsilon^{IJK} \psi^I_\alpha \psi^J_\beta \psi^K_\gamma 
\label{minop}
\end{equation}
which is not decorated with $\gamma$-matrices and $\alpha$, $\beta$ and 
$\gamma$ are spinor indices. There are several reasons for doing this. One is a
practical one to do with the size of the three loop calculation. There are 
$784$ three loop diagrams to determine and thus to keep computer run times to a
minimum it transpires that it is more efficient to evaluate the diagrams with 
(\ref{minop}) and then introduce the $\gamma$-matrix structure appropriate to 
each original operator of (\ref{opbas}) when summing the diagrams. In addition 
it also allows one to quickly construct that part of the mixing matrix relating
to the evanescent operators which are generated rather than have to repeat a 
full run. Indeed in that case the calculation would necessarily be slower as it
could involve $\Gamma_{(n)}^{\mu_1\ldots\mu_n}$ for values of $n$ up to $20$ 
for each of the three loop graphs. Another reason for proceeding with 
(\ref{minop}) is that we can extend the recent calculation of \cite{32} at the 
{\em same} time.

However, in choosing to calculate in this more general way there are several 
technical issues to be overcome which are rooted in the {\sc Mincer} algorithm.
It computes massless scalar $2$-point functions to three loops in dimensional 
regularization. With the $3$-quark operator zero momentum insertion in a quark 
$3$-point function the nullification of an external quark leg momentum ensures 
we have in effect a $2$-point function immediately. To obtain scalar integrals 
within the symbolic representation of each Feynman diagram we strip off the 
$\gamma$-matrix structure from the numerator as for (\ref{setup1}). An 
alternative approach would be to project out the Lorentz structure but this is 
too cumbersome especially as one has to have prior knowledge of the full 
structure of the Green's function at each loop order. Removing the 
$\gamma$-matrices instead leaves each diagram as a sum of Lorentz tensor 
integrals. Given that we are in the chiral limit these integrals will be of 
even rank. In the Feynman gauge they will be rank $(2l)$ where $l$ is the loop 
order. By contrast for a general linear covariant gauge these integrals will be
at most rank $(4l)$. In addition the {\sc Mincer} algorithm, \cite{7}, codes 
the internal loop momenta for each of the {\sc Mincer} topologies in terms of 
its own labelling. In particular each line of a topology is asigned an internal 
momentum label $p_i$. The conservation of energy momentum at each vertex is 
then encoded within the integration routine for that topology. In other words 
there are no {\sc Mincer} labels such as $(p_1-p_2)$ or $(Q-p_3-p_6)$ where $Q$
is the external momentum of the $2$-point function. Thus all the tensor 
integrals involve products of internal momentum vectors $p_i$ where 
$1$~$\leq$~$i$~$\leq$~$8$ for three loops. For lower loop orders there are
fewer internal momentum labels. Therefore, the problem of evaluating each
graph of the Green's function requires converting these even rank tensors into
scalar integrals, which are straightforward to compute in {\sc Mincer}, and
Lorentz tensors built from $\eta_{\mu\nu}$ and $Q_\mu$. The procedure for this
is straightforward. Using only knowledge of the rank we have written down the 
most general tensor basis for each rank and then determined the scalar
integral amplitude by the method of projection for each tensor. For instance,
for rank $2$, $4$ and $6$ there are respectively $2$, $10$ and $76$ tensors. 
Using {\sc Form} we have constructed the decomposition for an arbitrary 
numerator which is straightforward. For rank $8$ the projection matrix would be
$764$~$\times$~$764$. Rather than use a decomposition into $\eta_{\mu\nu}$ and 
$Q_\mu$ we chose the transverse and longitudinal projection tensors
\begin{equation}
P_{\mu\nu}(Q) ~=~ \eta_{\mu\nu} ~-~ \frac{Q_\mu Q_\nu}{Q^2} ~~~,~~~ 
L_{\mu\nu}(Q) ~=~ \frac{Q_\mu Q_\nu}{Q^2} 
\end{equation}
which satisfy the simple properties
\begin{eqnarray}
P_{\mu\nu}(Q) ~+~ L_{\mu\nu}(Q) &=& \eta_{\mu\nu} ~~,~~ \
P_{\mu\nu}(Q) P^{\nu\sigma}(Q) ~=~ P_\mu^{~\sigma}(Q) \nonumber \\
P_{\mu\nu}(Q) L^{\nu\sigma}(Q) &=& 0 ~~,~~ 
L_{\mu\nu}(Q) L^{\nu\sigma}(Q) ~=~ L_\mu^{~\sigma}(Q) ~. 
\end{eqnarray}
The benefit of this is that the full matrix becomes block diagonal with the
largest submatrix being $105$~$\times$~$105$. Once the general decomposition 
has been derived it is encoded as an integration module in {\sc Form} at the 
appropriate point in the overall algorithm. With rank $8$ decomposition the 
anomalous dimension matrix can be determined to two loops in an arbitrary 
linear covariant gauge. At this point we note that since the operators we 
consider are gauge invariant this allows us to check that in $\MSbar$ the two 
loop anomalous dimensions are independent of the gauge parameter. This is a 
strong check on the derivation of the {\sc Form} module encoding the mapping of
the Lorentz tensor integrals to the scalar integrals. Indeed the two loop 
result of \cite{20} was performed only in the Feynman gauge. For three loops 
one could in principle extend the construction to rank $12$. Instead we have 
taken a different approach and chosen to calculate the three loop diagrams 
solely in the Feynman gauge. In this case this only requires the decomposition 
up to rank $6$ which has already been tested for the arbitrary gauge 
calculation at two loops. Moreover, in the Feynman gauge the diagrams can be 
evaluated more quickly. The independent check on the three loop anomalous 
dimensions will be consistency with the renormalization group equation as the 
double and triple poles in $\epsilon$ are already fixed from the one and two 
loop renormalization constants. We believe this calculational approach from the
point of view of exploiting the properties of the gluon propagator in various 
gauges is the most efficient to deduce the anomalous dimensions. However, the 
finite part of this Green's function at three loops is not useful for lattice 
matching as the Landau gauge results will not be determined. Indeed it is not 
clear whether they would be meaningful anyway in this instance as the 
nullification of the momentum of an external quark leg could be problematic 
from the point of view of infrared divergences in the chiral limit. 

\sect{Three loop anomalous dimensions.}

While we have split the discussion on the technical aspects of each setup we
collect the results of our renormalization in one section. This is because for
both calculations we obtain the same results to two loops for an arbitrary 
linear covariant gauge. As indicated in the {\sc Mincer} discussion we have 
reasonable internal checks on the three loop part of the anomalous dimensions 
from the renormalization group equations. First, we record the naive anomalous
dimensions for the seed operators ${\cal O}_{(0)}$ and ${\cal O}_{(4)}$ where
the labels $1$ and $2$ refer to these operators respectively. To three loops in
$\MSbar$ we have  
\begin{eqnarray}
\tilde{\gamma}_{11}(a) &=& -~ 2 a - \left[ 2 \Nf + 51 \right] 
\frac{a^2}{9} \nonumber \\
&& + \left[ 260 \Nf^2 + [ 4320 \zeta(3) - 4656 ] \Nf + 1296 \zeta(3) + 23481 
\right] \frac{a^3}{162} ~+~ O(a^4) \nonumber \\
\tilde{\gamma}_{12}(a) &=& \frac{10}{3} a^2 
+ \left[ 216 \zeta(3) - 153 - 14 \Nf \right] \frac{a^3}{27} ~+~ O(a^4)
\nonumber \\
\tilde{\gamma}_{21}(a) &=& \frac{10}{3} a^2 
+ 4 \left[ 18 \zeta(3) + 331 - 22 \Nf \right] \frac{a^3}{9} ~+~ O(a^4)
\nonumber \\
\tilde{\gamma}_{22}(a) &=& -~ 2 a - \left[ 2 \Nf + 51 \right] 
\frac{a^2}{9} \nonumber \\ 
&& + \left[ 260 \Nf^2 + [ 4320 \zeta(3) + 1344 ] \Nf + 1296 \zeta(3) - 75519 
\right] \frac{a^3}{162} ~+~ O(a^4) 
\end{eqnarray}
where $\zeta(z)$ is the Riemann $\zeta$-function. The two loop parts are 
calculated for arbitrary $\alpha$ and agree with the full two loop mixing
matrix of \cite{20}. This is because at this order the $\gamma^5$ problem can
be ignored as was noted in \cite{20} and hence the two loop naive anomalous
dimensions are sufficient to determine the proton operator wave function 
renormalization. As we generate ${\cal O}_{(8)}$ or ${\cal O}_3$ we have to
include the next row and column of $\tilde{\gamma}_{ij}(a)$ to the appropriate
orders in $a$. Thus we have 
\begin{eqnarray}
\tilde{\gamma}_{33}(a) &=& -~ 2 a - \left[ 2 \Nf - 19869 \right] 
\frac{a^2}{9} ~+~ O(a^3) \nonumber \\
\tilde{\gamma}_{23}(a) &=& \frac{5}{864} a^2 
+ \left[ 4824 \zeta(3) - 2745 - 14 \Nf \right] \frac{a^3}{15552} ~+~ O(a^4)
\nonumber \\
\tilde{\gamma}_{31}(a) &=& \tilde{\gamma}_{32}(a) ~=~ O(a^3) \nonumber \\
\tilde{\gamma}_{13}(a) &=& O(a^4) ~. 
\end{eqnarray} 
The key terms in this are those in the upper triangle. In the formalism of
\cite{30} $\tilde{\gamma}_{31}(a)$ and $\tilde{\gamma}_{32}(a)$ correspond to 
the generalized $\beta$-functions used in the evanescent projection formalism 
to extract the effect the evanescent operators have on the strictly four 
dimensional anomalous dimensions, before we restore chiral symmetry and the 
anti-commutativity of $\gamma^5$. As these two anomalous dimensions are zero to
$O(a^3)$ then there can be no evanescent operator affect until {\em four} loops
in the full four dimensional anomalous dimensions.  
 
In order to complete the computation we require the finite renormalization
constants which restore the anti-commutativity of $\gamma^5$ in strictly four 
dimensions. As background it is worthwhile recalling the Larin procedure,
\cite{31}, for extracting the finite renormalization constant for the flavour 
non-singlet pseudo-scalar quark current, ${\cal O}_{m_5}$~$=$~$\bar{\psi} 
\gamma^5 \psi$. Here we assume the operator is inserted in a quark $2$-point 
function at zero momentum. With a naive anti-commuting $\gamma^5$ in 
$d$-dimensions one would simply anti-commute the $\gamma^5$ out of all the 
diagrams. However, retaining it inside diagrams and using the generalized 
$\Gamma_{(n)}$-matrices to compute the Green's function one obtains a finite 
Green's function in the four dimensional limit after the renormalization 
constants have been determined in the $\MSbar$ scheme. The value for this 
should be equivalent to that calculated with the naive anti-commuting 
$\gamma^5$ in $d$-dimensions. However, it is not the same and therefore to 
remove the discrepancy one defines a finite renormalization constant,
\cite{31}. If we define this as $Z^5$ for the pseudo-mass operator then more 
specifically the condition is defined by
\begin{equation} 
\left. \gamma^5 \langle \psi(p) {\cal O}_m(0) \bar{\psi}(-p) \rangle 
\right|_{d=4} ~=~ Z^5 \left. \langle \psi(p) {\cal O}_{m_5}(0) \bar{\psi}(-p) 
\rangle \right|_{d=4} 
\label{mass5}
\end{equation}
where ${\cal O}_m$~$=$~$\bar{\psi}\psi$ is the quark mass operator. While this
is the situation for a single operator with no mixing the renormalization of
the $3$-quark operator is complicated by the mixing. In this respect it is
notationally difficult to extend the definition of (\ref{mass5}) to the
matrix case. At a formal level the definition can be represented by
\begin{equation}
\langle {\cal O}_1 \rangle ~=~ Z^5 \odot \left( \gamma^5 \otimes \gamma^5
\right) \langle {\cal O}_2 \rangle
\label{barop5}
\end{equation} 
where the tensor product of the $\gamma^5$ matrices acts on the appropriate
spinor indices of the Green's function. Also $Z^5$ represents the finite
renormalization matrix and its effect within the finite part of the Green's
function which is the meaning of the $\odot$ multiplication. While this is the
normal Larin procedure where the naive $\MSbar$ anomalous dimensions are
computed first and then the finite renormalization, in practical terms it is
quicker to do both processes together. In other words we absorb a certain
finite part into our $\MSbar$ renormalization constants. The condition for this
is that the finite parts of both Green's functions after renormalization in
four dimensions are equivalent up to multiplying by 
$\gamma^5$~$\otimes$~$\gamma^5$. However, to assist others who wish to
reproduce the results we will {\em present} them as in the two stage Larin 
method. Thus in addition to the naive anomalous dimensions we have extracted 
the anomalous dimensions associated with the finite $\gamma^5$ renormalization
matrix. These are
\begin{eqnarray}
\gamma_{5,11}(a) &=& \gamma_{5,12}(a) ~=~ O(a^4) \nonumber \\
\gamma_{5,21}(a) &=& 125 \left[ 2 \Nf - 33 \right] \frac{a^3}{27} ~+~ O(a^4)
\nonumber \\
\gamma_{5,22}(a) &=& -~ 500 \left[ 2 \Nf - 33 \right] \frac{a^3}{27} ~+~ O(a^4) 
\label{anom5}
\end{eqnarray} 
where the linear factor in $\Nf$ derives from the one loop $\beta$-function for
$SU(3)$. In defining the finite renormalization condition (\ref{barop5}) we 
have not included the momentum configuration of the external quark legs. This 
is because (\ref{anom5}) has been derived for both setups. The emergence of the 
same finite renormalization from both calculations is an independent check. 
Moreover both computations were performed in an arbitrary linear covariant
gauge. However, as is evident from (\ref{anom5}) the result is independent
of $\alpha$. For all the quark bilinear current operators considered in 
\cite{31} and the diquark operators examined in \cite{40} which contained 
$\gamma^5$, the associated finite renormalization constant was also independent
of the gauge parameter. Indeed this follows from a simple observation that the
finite renormalization constant in the bilinear current cases is related via 
renormalization group arguments to the {\em naive} anomalous dimensions of the 
particular operators involved in the defining condition. In the case of our 
earlier example this would be the naive anomalous dimensions of ${\cal O}_m$ 
and ${\cal O}_{m_5}$. As the operators are gauge invariant and being 
renormalized in a mass independent scheme then as the operator anomalous 
dimensions are independent of the gauge parameter then so to is the finite 
renormalization constant. By contrast if the original operator renormalization 
had been performed in a mass dependent scheme then not only would the operator 
anomalous dimensions be gauge dependent but the finite renormalization constant
would too. This is of course barring accidental cancellations of the gauge 
parameter which is in principle possible. For the case of the operators 
${\cal O}_1$ and ${\cal O}_2$ a similar argument can be established. Though due
to the mixing it derives from within the formal renormalization condition 
(\ref{barop5}) and the specific way we have defined the finite renormalization 
constant within the explicit calculation of the Green's function of each 
operator in the one stage method of applying the Larin technique which we used
here. To summarize the upshot of that analysis translates into the equations
\begin{eqnarray}
\tilde{\gamma}_{11}(a) ~-~ \tilde{\gamma}_{22}(a) &=& \mu \frac{d~}{d\mu}
\ln Z^5_{22} \nonumber \\ 
\tilde{\gamma}_{12}(a) ~-~ \tilde{\gamma}_{21}(a) &=& \mu \frac{d~}{d\mu}
\ln Z^5_{21} ~.
\end{eqnarray}
These equations are analogous to the situation in the quark current example. As
each of the naive operator anomalous dimensions are gauge independent, since
the operators themselves are gauge invariant, and we are in the $\MSbar$ 
scheme, which is a mass independent scheme, then the two finite renormalization
constants are automatically also gauge independent. Indeed from the explicit 
expressions for the naive three loop anomalous dimensions it is straightforward
to check that (\ref{anom5}) are consistent with these general expressions. This
is a reassuring check since we derived (\ref{anom5}) from the finite part of 
(\ref{barop5}) which was calculated in an arbitrary linear covariant gauge in 
both calculational setups to two loops. It was only the three loop {\sc Mincer}
calculational which was carried out in the Feynman gauge. Therefore, there is 
consistency with that calculation too.

Equipped with this finite renormalization we can now determine the full four
dimensional mixing matrix using the naive anomalous dimensions. We find to 
three loops that 
\begin{eqnarray}
\gamma_{11}(a) ~=~ \gamma_{22}(a) &=& -~ 2 a - \left[ 2 \Nf + 51 \right] 
\frac{a^2}{9} \nonumber \\
&& + \left[ 260 \Nf^2 + [ 4320 \zeta(3) - 4656 ] \Nf + 1296 \zeta(3) + 23481 
\right] \frac{a^3}{162} \nonumber \\
&& +~ O(a^4) \nonumber \\
\gamma_{12}(a) ~=~ \gamma_{21}(a) &=& \frac{10}{3} a^2 
+ \left[ 216 \zeta(3) - 153 - 14 \Nf \right] \frac{a^3}{27} ~+~ O(a^4) ~.
\end{eqnarray} 
The effect of the finite renormalization has been to restore the symmetry of 
the mixing matrix so that the diagonal entries are equal and the off-diagonal
are the same but different to the other two. This structure was present at one
and two loops, \cite{1,2,3,20}. However, for the actual proton anomalous 
dimension we need to have the correct handedness of the up and down quarks
which requires the eigen-anomalous dimensions which are 
\begin{eqnarray}
\gamma_+(a) &=& -~ 2 a - \left[ 2 \Nf + 21 \right] \frac{a^2}{9}
+ \left[ 260 \Nf^2 + [ 4320 \zeta(3) - 4740 ] \Nf + 2592 \zeta(3) + 22563 
\right] \frac{a^3}{162} \nonumber \\
&& +~ O(a^4) \nonumber \\
\gamma_-(a) &=& -~ 2 a - \left[ 2 \Nf + 81 \right] \frac{a^2}{9}
+ \left[ 260 \Nf^2 + [ 4320 \zeta(3) - 4572 ] \Nf + 24399 \right] 
\frac{a^3}{162} \nonumber \\
&& +~ O(a^4) 
\label{anomdim}
\end{eqnarray}
where
\begin{equation}
\gamma_+(a) ~=~ \gamma_{11}(a) ~+~ \gamma_{12}(a) ~~,~~ 
\gamma_-(a) ~=~ \gamma_{11}(a) ~-~ \gamma_{12}(a) ~. 
\end{equation}
It is the latter, $\gamma_-(a)$, which corresponds to the proton. Numerically
we have 
\begin{eqnarray}
\gamma_+(a) &=& -~ 2.0000000 a - \left[ 0.2222222 \Nf + 2.3333333 \right] a^2
\nonumber \\
&& + \left[ 1.6049383 \Nf^2 + 2.7955915 \Nf + 158.5106882 \right] a^3 ~+~ 
O(a^4) \nonumber \\
\gamma_-(a) &=& -~ 2.0000000 a - \left[ 0.2222222 \Nf + 9.0000000 \right] a^2
\nonumber \\
&& + \left[ 1.6049383 \Nf^2 + 3.8326285 \Nf + 150.6111111 \right] a^3 ~+~ 
O(a^4) ~.
\end{eqnarray}
Having established the anomalous dimensions we can construct the 
renormalization group invariant current, $j_-$, using the same notation as
\cite{20}. It is defined in the conventional way by 
\begin{equation}
j_- ~=~ \exp \left( \int^a \frac{\gamma_-(z)}{\beta(z)} \, dz \right) 
\bar{j}_-(\mu)
\end{equation}
and satisfies
\begin{equation}
\mu \frac{dj_-}{d\mu} ~=~ 0 ~. 
\end{equation}
Solving
\begin{equation}
j_- ~\equiv~ \gamma_{j_-}(a) \bar{j}_-(\mu)
\end{equation}
explicitly we formally have 
\begin{eqnarray}
\gamma_{j_-}(a) &=& a^{\gamma_1/\beta_1} \left[ 1 + \left[ 
\frac{\gamma_2}{\beta_1} - \frac{\gamma_1\beta_2}{\beta_1^2} \right] a
\right. \nonumber \\
&& \left. ~~~~~~~~~
+ \left[ \frac{\gamma_3}{\beta_1} - \frac{\gamma_2\beta_2}{\beta_1^2}
- \frac{\gamma_1\beta_3}{\beta_1^2} + \frac{\gamma_2^2}{\beta_1^2}
+ \frac{\gamma_1\beta_2^2}{\beta_1^3}
- \frac{2\gamma_1\gamma_2\beta_2}{\beta_1^3}
+ \frac{\gamma_1^2\beta_2^2}{\beta_1^4} \right] \frac{a^2}{2} \right. 
\nonumber \\
&& \left. ~~~~~~~~~+ O(a^3) \right]
\end{eqnarray}
where 
\begin{eqnarray}
\gamma_-(a) &=& \gamma_1 a ~+~ \gamma_2 a^2 ~+~ \gamma_3 a^3 ~+~ O(a^4)
\nonumber \\
\beta(a) &=& \beta_1 a^2 ~+~ \beta_2 a^3 ~+~ \beta_3 a^4 ~+~ O(a^5) ~.
\end{eqnarray}
This produces 
\begin{eqnarray}
\gamma_{j_-}(a) &=& \left[ 1 - [4\Nf^2-588\Nf+2835] \frac{a}{3[2\Nf-33]^2} 
\right. \nonumber \\
&& \left. ~+ \left[ 2080 \Nf^5 + [ 34560 \zeta(3) - 157368 ] \Nf^4
+ [ 4596912 - 1710720 \zeta(3) ] \Nf^3 \right. \right. \nonumber \\
&& \left. \left. ~~~~~+ [ 28226880 \zeta(3) - 70113330 ] \Nf^2 
+ [ 580876920 - 155247840 \zeta(3) ] \Nf
\right. \right. \nonumber \\
&& \left. \left. ~~~~~- 1825381251 \right] \frac{a^2}{108[2\Nf-33]^4} ~+~ 
O(a^3) \right] a^{6/[33-2\Nf]} ~. 
\end{eqnarray}
For three flavours this gives
\begin{equation}
\left. \gamma_{j_-}(a) \right|_{\Nf=3} ~=~ \left[ 1 - \frac{41}{81} a
- \frac{[116640\zeta(3)+275215]}{26244} a^2 ~+~ O(a^3) \right] a^{2/9}
\end{equation}
or
\begin{equation}
\left. \gamma_{j_-}(a) \right|_{\Nf=3} ~=~ \left[ 1 - 0.5061728 a
- 15.8292531 a^2 ~+~ O(a^3) \right] a^{2/9}
\end{equation}
numerically. To gauge the effects of the two loop correction we can compare the
numerical value of $\left. \gamma_{j_-}(a) \right|_{\Nf=3}$ at one loop with
that at two loops for $\alpha_s$~$=$~$0.1$. We find that the two loop 
correction modifies the one loop value by around $0.1\%$.

\sect{Amplitudes}

In this section we record the explicit values of the Green's function
(\ref{setup1}) to two loops in an arbitrary linear covariant gauge in the 
$\MSbar$ scheme. This represents one of the main results of the article as it 
will be of use for lattice matching. We have\footnote{The full analytic form of
the amplitude for an arbitrary gauge and the anomalous dimensions, have been 
included in an attached electronic data file.}
\begin{eqnarray}
\left. \frac{}{} \left\langle \psi_\alpha(p) \psi_\beta(q) \psi_\gamma(-p-q) 
{\cal O}_{1\,\delta}(0) 
\right\rangle \right|_{\mbox{\footnotesize{symm}}} &=&
\left[ 1 + \frac{2}{9} \left[ 2 \pi^2 \alpha + 15 \alpha + 2 \pi^2 + 15 
\right. \right. \nonumber \\
&& \left. \left. ~~~~~~~~~- 3 \psi^\prime(\third) - 3 \psi^\prime(\third) 
\alpha \right] a \right. \nonumber \\
&& \left. + \left[ \left[ 2160 \psi^\prime(\third) - 1440 \pi^2 - 19008 
\right] \Nf 
+ 1440 \psi^\prime(\third)^2
\right. \right. \nonumber \\
&& \left. \left. ~~~
- \left[ 8262 \alpha^2 + 23652 \alpha + 122742 \right] \psi^\prime(\third)
\right. \right. \nonumber \\
&& \left. \left. ~~~
- 1920 \psi^\prime(\third) \pi^2
+ \left[ 711 \alpha + 207 \right] \psi^{\prime\prime\prime}(\third)
\right. \right. \nonumber \\
&& \left. \left. ~~~
+ \left[ 136080 \alpha - 392688 \right] s_2(\pisix)
\right. \right. \nonumber \\
&& \left. \left. ~~~
+ \left[ 785376 - 272160 \alpha \right] s_2(\pitwo)
\right. \right. \nonumber \\
&& \left. \left. ~~~
+ \left[ 654480 - 226800 \alpha \right] s_3(\pisix)
\right. \right. \nonumber \\
&& \left. \left. ~~~
+ \left[ 181440 \alpha - 523584 \right] s_3(\pitwo)
+ \left[ 88 - 1896 \alpha \right] \pi^4
\right. \right. \nonumber \\
&& \left. \left. ~~~
+ \left[ 5508 \alpha^2 + 15768 \alpha + 81828 \right] \pi^2
\right. \right. \nonumber \\
&& \left. \left. ~~~
- \left[ 18468 + 12636 \alpha \right] \Sigma
- \left[ 73224 \alpha + 23976 \right] \zeta(3) 
\right. \right. \nonumber \\
&& \left. \left. ~~~
+ 188244 \alpha + 403461 
+ \left[ 2929 - 1015 \alpha \right] \frac{\pi^3}{\sqrt{3}}
\right. \right. \nonumber \\
&& \left. \left. ~~~
+ \left[ 32724 - 11340 \alpha \right] \frac{\ln(3) \pi}{\sqrt{3}}
\right. \right. \nonumber \\
&& \left. \left. ~~~
+ \left[ 945 \alpha - 2727 \right] \frac{\ln^2(3) \pi}{\sqrt{3}}
\right] \frac{a^2}{2916} \right] I_{\alpha\delta} \otimes I_{\beta\gamma} 
\nonumber \\
&& +~ 5 \left[ \left[
648 \zeta(3) 
- 1323
- 540 \psi^\prime(\third) \pi^2
- 3888 s_2(\pisix)
\right. \right. \nonumber \\
&& \left. \left. ~~~~~~
+ 7776 s_2(\pitwo)
+ 6480 s_3(\pisix)
- 5184 s_3(\pitwo)
\right. \right. \nonumber \\
&& \left. \left. ~~~~~~
+ 360 \pi^2
+ 29 \frac{\pi^3}{\sqrt{3}}
+ 324 \frac{\ln(3) \pi}{\sqrt{3}}
\right. \right. \nonumber \\
&& \left. \left. ~~~~~~
- 27 \frac{\ln^2(3) \pi}{\sqrt{3}}
\right] \frac{a^2}{486} \right] \gamma^5_{\alpha\delta} \otimes 
\gamma^5_{\beta\gamma} \nonumber \\
&& +~ \left[ 4 \left[ 3 \psi^\prime(\third) \alpha 
+ 3 \psi^\prime(\third) - 2 \pi^2 \alpha - 2 \pi^2 \right] \frac{a}{81}
\right. \nonumber \\
&& \left. ~~~~+ \left[ \left[ 1056 \pi^2 - 1584 \psi^\prime(\third) 
\right] \Nf 
\right. \right. \nonumber \\
&& \left. \left. ~~~~~~~~ 
- \left[ 333 \alpha + 513 \right] \psi^{\prime\prime\prime}(\third)
\right. \right. \nonumber \\
&& \left. \left. ~~~~~~~~ 
+ \left[ 2322 \alpha^2 - 1188 \alpha - 1350 \right] \psi^\prime(\third)
\right. \right. \nonumber \\
&& \left. \left. ~~~~~~~~
- \left[ 104976 \alpha + 295488 \right] s_2(\pisix)
\right. \right. \nonumber \\
&& \left. \left. ~~~~~~~~
+ \left[ 209952 \alpha + 590976 \right] s_2(\pitwo)
\right. \right. \nonumber \\
&& \left. \left. ~~~~~~~~
+ \left[ 174960 \alpha + 492480 \right] s_3(\pisix)
\right. \right. \nonumber \\
&& \left. \left. ~~~~~~~~
- \left[ 139968 \alpha + 393984 \right] s_3(\pitwo)
\right. \right. \nonumber \\
&& \left. \left. ~~~~~~~~
+ \left[ 888 \alpha + 1368 \right] \pi^4
\right. \right. \nonumber \\
&& \left. \left. ~~~~~~~~
+ \left[ 900 + 792 \alpha - 1548 \alpha^2 \right] \! \pi^2
\right. \right. \nonumber \\
&& \left. \left. ~~~~~~~~
+ \left[ 4212 \alpha + 4212 \right] \Sigma
+ \left[ 15552 \alpha - 5184 \right] \zeta(3)
\right. \right. \nonumber \\
&& \left. \left. ~~~~~~~~
+ \left[ 2204 + 783 \alpha \right] \frac{\pi^3}{\sqrt{3}}
\right. \right. \nonumber \\
&& \left. \left. ~~~~~~~~
+ \left[ 24624 + 8748 \alpha \right] \frac{\ln(3) \pi}{\sqrt{3}}
\right. \right. \nonumber \\
&& \left. \left. ~~~~~~~~
- \left[ 729 \alpha + 2052 \right] \frac{\ln^2(3) \pi}{\sqrt{3}}
\right] \frac{a^2}{4374} \right] \frac{I_{\alpha\delta} \otimes 
\sigma^{pq}_{\beta\gamma}}{\mu^2} 
\nonumber \\
&& +~ 5 \left[ \left[
648 \zeta(3) 
+ 216 \psi^\prime(\third) \pi^2
- 18 \psi^{\prime\prime\prime}(\third)
\right. \right. \nonumber \\
&& \left. \left. ~~~~~~
- 3888 s_2(\pisix)
+ 7776 s_2(\pitwo)
+ 6480 s_3(\pisix)
\right. \right. \nonumber \\
&& \left. \left. ~~~~~~
- 5184 s_3(\pitwo)
+ 48 \pi^4
- 144 \pi^2
+ 29 \frac{\pi^3}{\sqrt{3}}
\right. \right. \nonumber \\
&& \left. \left. ~~~~~~
+ 324 \frac{\ln(3) \pi}{\sqrt{3}}
\right. \right. \nonumber \\
&& \left. \left. ~~~~~~
- 27 \frac{\ln^2(3) \pi}{\sqrt{3}}
\right] \frac{a^2}{4374} \right] 
\frac{\gamma^5_{\alpha\delta} \otimes \left(\gamma^5 
\sigma^{pq}\right)_{\beta\gamma}}{\mu^2} \nonumber \\
&& +~ \left[ 2 \left[ 3 \psi^\prime(\third) \alpha + 3 \psi^\prime(\third) 
- 2 \pi^2 \alpha - 2 \pi^2 \right] \frac{a}{81}
\right. \nonumber \\
&& \left. ~~~~+ \left[ \left[ 912 \pi^2 - 1368 \psi^\prime(\third) \right] \Nf
\right. \right. \nonumber \\
&& \left. \left. ~~~~~~~~ 
- \left[ 333 \alpha + 162 \right] \psi^{\prime\prime\prime}(\third)
\right. \right. \nonumber \\
&& \left. \left. ~~~~~~~~ 
+ \left[ 2322 \alpha^2 - 1188 \alpha + 31050 \right] \psi^\prime(\third)
\right. \right. \nonumber \\
&& \left. \left. ~~~~~~~~
+ \left[ 19440 - 104976 \alpha \right] s_2(\pisix)
\right. \right. \nonumber \\
&& \left. \left. ~~~~~~~~
+ \left[ 209952 \alpha - 38880 \right] s_2(\pitwo)
\right. \right. \nonumber \\
&& \left. \left. ~~~~~~~~
+ \left[ 174960 \alpha - 32400 \right] s_3(\pisix)
\right. \right. \nonumber \\
&& \left. \left. ~~~~~~~~
+ \left[ 25920 - 139968 \alpha \right] s_3(\pitwo)
\right. \right. \nonumber \\
&& \left. \left. ~~~~~~~~
+ \left[ 888 \alpha + 432 \right] \pi^4
+ \left[ 4212 \alpha + 7128 \right] \Sigma
\right. \right. \nonumber \\
&& \left. \left. ~~~~~~~~
- \left[ 20700 - 792 \alpha + 1548 \alpha^2 \right] \! \pi^2
\right. \right. \nonumber \\
&& \left. \left. ~~~~~~~~
+ \left[ 15552 \alpha - 19764 \right] \zeta(3)
\right. \right. \nonumber \\
&& \left. \left. ~~~~~~~~
+ \left[ 8748 \alpha - 1620 \right] \frac{\ln(3) \pi}{\sqrt{3}}
\right. \right. \nonumber \\
&& \left. \left. ~~~~~~~~
+ \left[ 135 - 729 \alpha \right] \frac{\ln^2(3) \pi}{\sqrt{3}}
\right. \right. \nonumber \\
&& \left. \left. ~~~~~~~~
+ \left[ 783 \alpha - 145 \right] \frac{\pi^3}{\sqrt{3}}
\right] \frac{a^2}{8748} \right] 
\frac{\sigma^{p\mu}_{\alpha\delta} \otimes \sigma_{p\mu \, \beta\gamma}}{\mu^2} 
\nonumber \\
&& +~ \left[ 4 \left[ 3 \psi^\prime(\third) \alpha 
+ 3 \psi^\prime(\third) - 2 \pi^2 \alpha - 2 \pi^2 \right] \frac{a}{81}
\right. \nonumber \\
&& \left. ~~~~+ \left[ \left[ 1968 \pi^2 - 2952 \psi^\prime(\third) \right] \Nf
\right. \right. \nonumber \\
&& \left. \left. ~~~~~~~~ 
- \left[ 666 \alpha + 765 \right] \psi^{\prime\prime\prime}(\third)
\right. \right. \nonumber \\
&& \left. \left. ~~~~~~~~ 
+ \left[ 4644 \alpha^2 - 2376 \alpha + 30780 \right] \psi^\prime(\third)
\right. \right. \nonumber \\
&& \left. \left. ~~~~~~~~
- \left[ 295488 + 209952 \alpha \right] s_2(\pisix)
\right. \right. \nonumber \\
&& \left. \left. ~~~~~~~~
+ \left[ 419904 \alpha + 590976 \right] s_2(\pitwo)
\right. \right. \nonumber \\
&& \left. \left. ~~~~~~~~
+ \left[ 349920 \alpha + 492480 \right] s_3(\pisix)
\right. \right. \nonumber \\
&& \left. \left. ~~~~~~~~
- \left[ 393984 + 279936 \alpha \right] s_3(\pitwo)
\right. \right. \nonumber \\
&& \left. \left. ~~~~~~~~
+ \left[ 1776 \alpha + 2040 \right] \pi^4
+ \left[ 8424 \alpha + 11340 \right] \Sigma
\right. \right. \nonumber \\
&& \left. \left. ~~~~~~~~
- \left[ 20520 - 1584 \alpha + 3096 \alpha^2 \right] \! \pi^2
\right. \right. \nonumber \\
&& \left. \left. ~~~~~~~~
+ \left[ 31104 \alpha - 21708 \right] \zeta(3)
\right. \right. \nonumber \\
&& \left. \left. ~~~~~~~~
+ \left[ 17496 \alpha + 24624 \right] \frac{\ln(3) \pi}{\sqrt{3}}
\right. \right. \nonumber \\
&& \left. \left. ~~~~~~~~
- \left[ 2052 + 1458 \alpha \right] \frac{\ln^2(3) \pi}{\sqrt{3}}
\right. \right. \nonumber \\
&& \left. \left. ~~~~~~~~
+ \left[ 2204 + 1566 \alpha \right] \frac{\pi^3}{\sqrt{3}}
\right] \frac{a^2}{8748} \right] 
\frac{\sigma^{p\mu}_{\alpha\delta} \otimes \sigma_{q\mu \, \beta\gamma}}{\mu^2} 
\nonumber \\
&& +~ \left[ \left[ \left[ 212 \psi^\prime(\third) - 144 \pi^2 \right] \Nf
+ 441 \psi^{\prime\prime\prime}(\third)
\right. \right. \nonumber \\
&& \left. \left. ~~~~~~
+ 31320 \psi^\prime(\third)
+ 334368 s_2(\pisix)
- 668736 s_2(\pitwo)
\right. \right. \nonumber \\
&& \left. \left. ~~~~~~
- 557280 s_3(\pisix)
+ 445824 s_3(\pitwo)
- 1176 \pi^4
\right. \right. \nonumber \\
&& \left. \left. ~~~~~~
- 20880 \pi^2
+ 2916 \Sigma
- 17820 \zeta(3)
\right. \right. \nonumber \\
&& \left. \left. ~~~~~~
- 2494 \frac{\pi^3}{\sqrt{3}}
- 27864 \frac{\ln(3) \pi}{\sqrt{3}}
\right. \right. \nonumber \\
&& \left. \left. ~~~~~~
+ 2322 \frac{\ln^2(3) \pi}{\sqrt{3}}
\right] \frac{a^2}{8748} \right] 
\frac{\sigma^{q\mu}_{\alpha\delta} \otimes 
\sigma_{p\mu \, \beta\gamma}}{\mu^2} \nonumber \\
&& +~ O(a^3) 
\label{op1amps}
\end{eqnarray}
where $\mbox{symm}$ denotes (\ref{symmpt}) and $\psi(z)$ is the derivative of 
the logarithm of the Euler $\Gamma$-function. Other various quantities are 
defined by
\begin{equation}
\Sigma ~=~ {\cal H}^{(2)}_{31} ~+~ {\cal H}^{(2)}_{43} ~~,~~
s_n(z) ~=~ \frac{1}{\sqrt{3}} \Im \left[ \mbox{Li}_n \left(
\frac{e^{iz}}{\sqrt{3}} \right) \right]
\end{equation}
where $\mbox{Li}_n(z)$ is the polylogarithm function. We retain the notation of
\cite{13} here in defining the quantity $\Sigma$ which is a linear combination 
of two harmonic polylogarithms, ${\cal H}^{(2)}_{31}$ and 
${\cal H}^{(2)}_{43}$. More background to their appearance in the basic master 
integrals can be found in Appendix A of \cite{13}. The theory for such harmonic
polylogarithms was developed in \cite{50}. These polylogarithms together with 
the other quantities such as $\zeta(3)$ and $\ln(3)$ emerge from the one and 
two loop master integrals, \cite{46,47,48,49}. Recently, the mathematics of 
these masters has been studied in the context of cyclotomic polynomials and 
harmonic polylogarithms in \cite{51}. There an insight has been given for which 
particular polylogarithms and other such numbers will arise in the higher loop 
order master integrals. We also note that our convention in (\ref{op1amps}) is 
that when a Lorentz index is contracted with one of the external momenta $p$ or
$q$ then the Lorentz index is replaced by the momentum to compactify notation. 

We only include the expression for the Green's function containing ${\cal O}_1$
since that for ${\cal O}_2$ can be readily deduced by multiplying 
(\ref{op1amps}) by $\gamma^5$~$\otimes$~$\gamma^5$. Then the sum and difference
of these two cases will give respectively the finite parts of the Green's 
function of the eigen-operators of $\gamma_{ij}(a)$. However, we have checked 
that both are in agreement in four dimensions. This is because while we have 
derived the finite renormalization required to restore anti-commutativity of 
$\gamma^5$ in four dimensions that was essentially based on the two structures 
$I$~$\otimes$~$I$ and $\gamma^5$~$\otimes$~$\gamma^5$. Aside from the fact that
the $3$-quark operators are renormalizable, the reason for this is that these 
channels ordinarily contain the divergences in $\epsilon$. Therefore in keeping
with the Larin method, \cite{31}, they are used to define the finite 
renormalization matrix. In choosing this procedure it transpires that the 
coefficients in the other channels are in agreement when the conventions on the
charge conjugation matrix, $C$, are respected. Next in extracting 
(\ref{op1amps}) from our $d$-dimensional expression in addition to 
(\ref{gammap}) we have to be careful in taking the four dimensional limit of 
generalized $\Gamma$-matrices where there are contractions with either or both 
of the external momenta $p$ and $q$. So in addition to (\ref{gammap}) we have 
used
\begin{eqnarray}
\left. \Gamma_{(4)}^{p\mu\nu\sigma} \otimes \Gamma_{(4) \, q\mu\nu\sigma}
\right|_{d=4} &=& 6 pq \gamma^5 \otimes \gamma^5 \nonumber \\ 
\left. \Gamma_{(4)}^{pq\mu\nu} \otimes \Gamma_{(4) \, pq\mu\nu}
\right|_{d=4} &=& 2 p^2 q^2 \gamma^5 \otimes \gamma^5 ~-~ 2 (pq)^2 \gamma^5 
\otimes \gamma^5 
\end{eqnarray}
for (\ref{setup1}). The explicit values, (\ref{symmpt}) and (\ref{symmptpq}),
can be substituted in these general expressions. We also should comment on the 
structure of (\ref{op1amps}) in the various channels. One approach to extract 
the finite part might have been to project the coefficients of each Lorentz 
structure. While we did not follow that line one can observe from the final 
result that there are seven such channels at two loops in {\em four} 
dimensions. In $d$-dimensions, prior to taking the limit to four dimensions 
after renormalization, there are more than seven channels. Moreover, it seems 
that not all possible structures are generated and it would appear that that 
strategy would require more effort than is necessary. Also internally it would 
require the manipulation of long strings of $\gamma$-matrices after applying 
the transpose of a set of these matrices due to (\ref{Cgam}). This would slow 
the symbolic manipulation programmes.  

While (\ref{op1amps}) represents the evaluation of the Green's function to two
loops in $\MSbar$ analytically, for practical purposes the numerical value is
more useful. Therefore, we have evaluated (\ref{op1amps}) to seven decimal
places and find 
\begin{eqnarray}
\left. \frac{}{} \left\langle \psi_\alpha(p) \psi_\beta(q) \psi_\gamma(-p-q) 
{\cal O}_{1\,\delta}(0) 
\right\rangle \right|_{\mbox{\footnotesize{symm}}} &=&
\left[ 1 + \left[ 0.9894261 \alpha + 0.9894261 \right] a \right. \nonumber \\
&& \left. + \left[ 41.5310566 + 6.7082190 \alpha + 2.8995053 \alpha^2 
\right. \right. \nonumber \\
&& \left. \left. ~~~- 3.9141771 \Nf \right] a^2 \right] I_{\alpha\delta} 
\otimes I_{\beta\gamma} \nonumber \\
&& -~ 1.6908864 a^2 \gamma^5_{\alpha\delta} \otimes \gamma^5_{\beta\gamma} 
\nonumber \\
&& + \left[ \left[ 0.5208683 \alpha + 0.5208683 \right] a \right. \nonumber \\
&& \left. ~~~+ \left[ 16.3956216 + 6.1021902 \alpha + 1.8664447 \alpha^2 
\right. \right. \nonumber \\
&& \left. \left. ~~~~~~~- 1.2732336 \Nf \right] a^2 \right] 
\frac{I_{\alpha\delta} \otimes \sigma^{pq}_{\beta\gamma}}{\mu^2} \nonumber \\
&& -~ 0.3372607 a^2 \frac{\gamma^5_{\alpha\delta} \otimes \left( \gamma^5 
\sigma^{pq}\right)_{\beta\gamma}}{\mu^2} \nonumber \\
&& + \left[ \left[ 0.2604341 \alpha + 0.2604341 \right] a \right. \nonumber \\
&& \left. ~~~+ \left[ 9.4012326 + 3.0510951 \alpha + 0.9332223 \alpha^2
\right. \right. \nonumber \\
&& \left. \left. ~~~~~~~- 0.5498054 \Nf \right] a^2 \right]
\frac{\sigma^{p\mu}_{\alpha\delta} \otimes \sigma_{p\mu \, \beta\gamma}}{\mu^2}
\nonumber \\
&& + \left[ \left[ 0.5208683 \alpha + 0.5208683 \right] a \right. \nonumber \\
&& \left. ~~~+ \left[ 17.4304130 + 6.1021902 \alpha + 1.8664447 \alpha^2
\right. \right. \nonumber \\
&& \left. \left. ~~~~~~~- 1.1864222 \Nf \right] a^2 \right] 
\frac{\sigma^{p\mu}_{\alpha\delta} \otimes \sigma_{q\mu \, \beta\gamma}}{\mu^2}
\nonumber \\
&& + \left[ 1.3720521 + 0.0868114 \Nf \right] a^2
\frac{\sigma^{q\mu}_{\alpha\delta} \otimes \sigma_{p\mu \, \beta\gamma}}{\mu^2}
\nonumber \\
&& +~ O(a^3)
\end{eqnarray}
where we have used 
\begin{eqnarray}
\zeta(3) &=& 1.20205690 ~~,~~ \Sigma ~=~ 6.34517334 ~~,~~
\psi^\prime\left( \frac{1}{3} \right) ~=~ 10.09559713 ~, \nonumber \\
\psi^{\prime\prime\prime}\left( \frac{1}{3} \right) &=& 488.1838167 ~~,~~
s_2\left( \frac{\pi}{2} \right) ~=~ 0.32225882 ~~,~~
s_2\left( \frac{\pi}{6} \right) ~=~ 0.22459602 ~, \nonumber \\
s_3\left( \frac{\pi}{2} \right) &=& 0.32948320 ~~,~~
s_3\left( \frac{\pi}{6} \right) ~=~ 0.19259341 
\end{eqnarray}
as the input values in this exercise. Finally, we have not included the finite 
part of the Green's function (\ref{setup2}) to two or three loops since lattice
measurements would require a zero momentum quark. This is a very difficult task
numerically on the lattice. Moreover, only the two loop results of 
(\ref{setup1}) are in an arbitrary linear covariant gauge since we restricted 
the three loop {\sc Mincer} calculation to the Feynman gauge.

\sect{General operator}

We now consider a generalization of the basic spin $\half$ operators we have
focused on so far. Recently, Kr\"{a}nkl and Manashov, \cite{32}, have 
introduced the operator
\begin{equation}
{\cal O}^{ijk}_{\alpha\beta\gamma} ~=~ \epsilon^{IJK} \psi^{iI}_\alpha
\psi^{jJ}_\beta \psi^{kK}_\gamma
\label{genop}
\end{equation}
which has no contractions over the spinor indices and $i$, $j$ and $k$ are
flavour indices. By considering the renormalization of this basic operator and 
its mixing into operators of the same dimension they managed to derive two loop
expressions for several other operator aside from the two considered in the
previous sections. However, the anomalous dimensions recorded in \cite{32} are 
not in the $\MSbar$ scheme. Despite this we have extended the results of 
\cite{32} to three loops. This is straightforward as the symbolic manipulation 
programmes used to derive the $\MSbar$ renormalization of the $(\half,0)$ 
operators was sufficiently general and hence adaptable to (\ref{genop}). First,
we recall the notation and formalism for the renormalization of (\ref{genop}), 
\cite{32}. The bare operator mixes into an infinite set of related operators 
which involve the generalized $\gamma$-matrices, 
$\Gamma_{(n)}^{\mu_1\ldots\mu_n}$. Though at each order in perturbation theory 
the number of generated operators is finite. Specifically,
\begin{equation}
{\cal O}^{ijk}_{\mbox{\footnotesize o} \, \alpha\beta\gamma} ~=~ 
Z_{\alpha ~ \beta ~~ \gamma}^{~\alpha^\prime ~ \beta^\prime \, \gamma^\prime}
{\cal O}^{ijk}_{\alpha^\prime\beta^\prime\gamma^\prime} 
\end{equation}
where the renormalization constant matrix
$Z_{\alpha ~ \beta ~~ \gamma}^{~\alpha^\prime ~ \beta^\prime \, \gamma^\prime}$
is given by
\begin{equation}
Z_{\alpha\alpha^\prime\beta\beta^\prime\gamma\gamma^\prime} ~=~
\delta_{\alpha\alpha^\prime} \delta_{\beta\beta^\prime}
\delta_{\gamma\gamma^\prime} ~+~ \sum_{k} a_{mnp}(\epsilon) \, 
\Gg_{mnp (\alpha\alpha^\prime | \beta\beta^\prime | \gamma\gamma^\prime)} ~.
\end{equation}
Here the poles in $\epsilon$ are contained within the function 
$a_{mnp}(\epsilon)$ 
where $k$~$\equiv$~$mnp$ is a label which indicates the basic 
$\Gamma_{(n)}$-matrix structure and there is no sum over individual $m$, $n$ 
and $p$ but over the corporate label $k$, and
\begin{equation}
a_{mnp}(\epsilon) ~=~ \sum_{n=1}^\infty \frac{a^{(n)}_{mnp}}{\epsilon^n} ~.
\end{equation}
As there are three open spinor indices in (\ref{genop}) the 
$\Gamma_{(n)}$-matrix structure is of the form, \cite{32},
\begin{equation}
\Gg_{mnp ( \alpha\alpha^\prime | \beta\beta^\prime | \gamma\gamma^\prime)} ~=~
\Gamma_{(m) \alpha\alpha^\prime} \otimes \Gamma_{(n) \beta\beta^\prime} \otimes
\Gamma_{(p) \gamma\gamma^\prime} ~.
\end{equation}
We will omit spinor indices from this point and use tensor product notation as
it is clearer. We have not included the Lorentz indices here but the 
contractions are across different $\Gamma$-matrices due to the antisymmetric 
property and there are no free Lorentz indices. It turns out that from explicit
calculations the $\Gamma_{(n)}$-structures appear in a symmetric form. To three
loops, using the same notation as \cite{32}, these are 
\begin{eqnarray}
\Cc_0 &=& \Gamma_{000} ~~,~~ 
\Cc_2 ~=~ \Gamma_{220} ~+~ \Gamma_{202} ~+~ \Gamma_{022} ~~,~~ 
\Cc_4 ~=~ \Gamma_{440} ~+~ \Gamma_{404} ~+~ \Gamma_{044} \nonumber \\
\Cc_6 &=& \Gamma_{660} ~+~ \Gamma_{606} ~+~ \Gamma_{066} ~~,~~ 
\Cc_{222} ~=~ \Gamma_{222} ~~,~~ 
\Cc_{42} ~=~ \Gamma_{422} ~+~ \Gamma_{242} ~+~ \Gamma_{224} \nonumber \\
\Cc_{442} &=& \Gamma_{442} ~+~ \Gamma_{424} ~+~ \Gamma_{244} ~~,~~ 
\Cc_{444} ~=~ \Gamma_{444} \nonumber \\
\Cc_{642} &=& \Gamma_{642} ~+~ \Gamma_{624} ~+~ \Gamma_{462} ~+~ \Gamma_{426} 
~+~ \Gamma_{246} ~+~ \Gamma_{264} 
\label{Ccdef}
\end{eqnarray}
where $\Cc_i$ retain the same spinor index structure as $\Gg$. We have included
$\Cc_{222}$ and $\Cc_{442}$ in this list as they appear at intermediate parts
of the renormalization but are absent in the final expression in keeping with
the expectation that the total number of $\gamma$-matrices should be divisible
by four. To avoid any confusion the explicit contraction of the Lorentz indices
in each definition of (\ref{Ccdef}) is
\begin{eqnarray}
\Gamma_{000} &=& \Gamma_{(0)} \otimes \Gamma_{(0)} \otimes \Gamma_{(0)}
\nonumber \\
\Gamma_{220} &=& \Gamma_{(2)}^{\mu_1\mu_2} \otimes 
\Gamma_{(2) \, \mu_1\mu_2} \otimes \Gamma_{(0)}
\nonumber \\
\Gamma_{440} &=& \Gamma_{(4)}^{\mu_1\mu_2\mu_3\mu_4} \otimes 
\Gamma_{(4) \, \mu_1\mu_2\mu_3\mu_4} \otimes \Gamma_{(0)}
\nonumber \\
\Gamma_{660} &=& \Gamma_{(6)}^{\mu_1\mu_2\mu_3\mu_4\mu_5\mu_6} \otimes 
\Gamma_{(6) \, \mu_1\mu_2\mu_3\mu_4\mu_5\mu_6} \otimes \Gamma_{(0)}
\nonumber \\
\Gamma_{222} &=& \Gamma_{(2)}^{\mu_1\mu_2} \otimes 
\Gamma_{(2) \, \mu_1\mu_3} \otimes \Gamma_{(2) \, \mu_2}^{~~~~~~\mu_3} 
\nonumber \\
\Gamma_{422} &=& \Gamma_{(4)}^{\mu_1\mu_2\mu_3\mu_4} \otimes 
\Gamma_{(2) \, \mu_1\mu_2} \otimes \Gamma_{(2) \, \mu_3\mu_4} 
\nonumber \\
\Gamma_{442} &=& \Gamma_{(4)}^{\mu_1\mu_2\mu_3\mu_4} \otimes 
\Gamma_{(4) \, \mu_1\mu_2\mu_3}^{~~~~~~~~~~~\mu_5} \otimes 
\Gamma_{(2) \, \mu_4\mu_5} 
\nonumber \\
\Gamma_{444} &=& \Gamma_{(4)}^{\mu_1\mu_2\mu_3\mu_4} \otimes 
\Gamma_{(4) \, \mu_1\mu_2}^{~~~~~~~~~\mu_5\mu_6} \otimes 
\Gamma_{(4) \, \mu_3\mu_4\mu_5\mu_6} 
\nonumber \\
\Gamma_{642} &=& \Gamma_{(6)}^{\mu_1\mu_2\mu_3\mu_4\mu_5\mu_6} \otimes 
\Gamma_{(4) \, \mu_1\mu_2\mu_3\mu_4} \otimes 
\Gamma_{(2) \, \mu_5\mu_6} 
\label{Gdef}
\end{eqnarray}
with the obvious permutation of $m$, $n$ and $p$ to define the forms in
(\ref{Ccdef}) not listed in (\ref{Gdef}). In (\ref{Ccdef}) $\Cc_0$, $\Cc_2$, 
$\Cc_4$ and $\Cc_{42}$ arise at two loops, \cite{32}, and $\Cc_6$, $\Cc_{222}$, 
$\Cc_{442}$, $\Cc_{444}$ and $\Cc_{642}$ only appear at three loops. That there
are no other structures to this order is elementary to deduce from the fact 
that beginning with (\ref{genop}) there are $4$, $8$ and $12$ possible 
$\gamma$-matrices in each of the respective one, two and three loop Feynman 
diagrams. With the absence of free Lorentz indices and the antisymmetry 
(\ref{Ccdef}) are all that survive.

In \cite{32} in order to ease the derivation of the anomalous dimension of 
(\ref{genop}) at two loops in four dimensions a relation was derived for 
$\Cc_{42}$ in $d$-dimensions which was  
\begin{equation}
\Cc_{42} ~=~ -~ 3 d(d-1) \Cc_0 ~-~ 2 (d-3) \Cc_2 ~-~ \frac{1}{2} \Cc_4 ~+~ 
\frac{1}{2} \Cc_2^2 
\label{Ccreln1}
\end{equation} 
where the product of the $\Cc_l$ is regarded as the multiplication of the 
constituent $\gamma$-matrices. While this is a relation in $d$-dimensions 
ultimately we will require the anomalous dimension in four dimensions and as 
noted in \cite{32} then 
\begin{equation}
\left. \frac{}{} \Cc_4 \right|_{d=4} ~=~ 24 \left[ 
\gamma^5 \otimes \gamma^5 \otimes I ~+~
\gamma^5 \otimes I \otimes \gamma^5 ~+~
I \otimes \gamma^5 \otimes \gamma^5 \right] ~. 
\end{equation}
In addition we have the similar but more trivial relations
\begin{eqnarray}
\left. \frac{}{} \Cc_0 \right|_{d=4} &=&  I \otimes I \otimes I \nonumber \\
\left. \frac{}{} \Cc_2 \right|_{d=4} &=& \left[ 
\sigma^{\mu\nu} \otimes \sigma_{\mu\nu} \otimes I ~+~
\sigma^{\mu\nu} \otimes I \otimes \sigma_{\mu\nu} ~+~
I \otimes \sigma^{\mu\nu} \otimes \sigma_{\mu\nu} \right] ~. 
\end{eqnarray}
At three loops two new structures emerge, $\Cc_6$ and $\Cc_{642}$, which 
involve the evanescent $\Gamma_{(6)}^{\mu_1\ldots\mu_6}$ matrix. However, 
similar to (\ref{Ccreln1}) one can deduce that in $d$-dimensions 
\begin{eqnarray}
\Cc_6 &=& -~ 12 d(d-1)(2d-1) \Cc_0 ~-~ 3 (d-1)(7d-24) \Cc_2 ~-~ 6 (2d-5) \Cc_4 
\nonumber \\
&& +~ 2 (3d-4) \Cc_2^2 ~-~ \frac{1}{2} \Cc_2^3 ~+~ \frac{3}{2} \Cc_2 \Cc_4 ~+~ 
3 \Cc_{444} \nonumber \\
\Cc_{642} &=& 12 d(d-1)(2d-7) \Cc_0 ~+~ 9 (d^2-9d+16) \Cc_2 ~+~ 2 (2d-5) \Cc_4 
\nonumber \\ 
&& -~ 2 (3d-10) \Cc_2^2 ~+~ \frac{1}{2} \Cc_2^3 ~-~ 
\frac{1}{2} \Cc_2 \Cc_4 ~-~ 3 \Cc_{444} ~. 
\label{evmatrel}
\end{eqnarray} 
So these evanescent combinations can be expressed in terms of $\Cc_i$ which do
not involve any $\Gamma_{(n)}^{\mu_1\ldots\mu_n}$ with $n$~$\geq$~$5$. For the
restriction to four dimensions we have the additional relation 
\begin{equation}
\left. \frac{}{} \Cc_{444} \right|_{d=4} ~=~ 0 ~.
\end{equation}
This follows trivially from the antisymmetry property. Given the presence of
$\Gamma_{(4)}^{\mu_1\mu_2\mu_3\mu_4}$ one possibility for this could have been 
$\gamma^5 \otimes \gamma^5 \otimes \gamma^5$. It is easy to see that this is 
excluded when one examines the pattern of Lorentz indices in four dimensions. 
The use of the relations (\ref{evmatrel}) can be viewed within the approach of 
\cite{32} as a variation of the formalism of \cite{30} which was introduced to 
include the effect of evanescent operators in the renormalization group 
functions. 

Equipped with these identities we have extracted the three loop anomalous
dimensions (\ref{genop}) in the $\MSbar$ scheme from the same three loop 
{\sc Mincer} computation as in previous sections. Though in this case we do not
contract the free spinor indices to produce a spin $\half$ operator. The full
result is  
\begin{eqnarray}
\gamma_{\cal O}(a) &=& -~ \frac{1}{6} \Cc_2 a + \left[ [ 36 - 2\Nf ] \Cc_0
+ \left[ \frac{1}{54} \Nf - \frac{47}{36} \right] \Cc_2 - \frac{1}{72} \Cc_2^2 
+ \frac{5}{72} \Cc_4 \right] a^2 \nonumber \\
&& + \left[ \left[ \frac{10}{9} \Nf^2 - \frac{853}{9} \Nf + \frac{8047}{9}
+ 17 \zeta(3) \right] \Cc_0 
\right. \nonumber \\
&& \left. ~~~~
+ \left[ \frac{433}{36} \zeta(3) - \frac{5873}{216} 
+ \left[ \frac{71}{54} + \frac{20}{9} \zeta(3) \right] \Nf 
+ \frac{13}{162} \Nf^2 \right] \Cc_2 
\right. \nonumber \\
&& \left. ~~~~
+ \left[ \frac{209}{648} - \frac{71}{54} \zeta(3) - \frac{1}{648} \Nf 
\right] \Cc_2^2
+ \left[ \frac{1}{54} \zeta(3) - \frac{5}{1296} \right] \Cc_2^3
\right. \nonumber \\
&& \left. ~~~~
+ \left[ \frac{29}{24} \zeta(3) - \frac{91}{144} - \frac{7}{648} \Nf 
\right] \Cc_4
+ \left[ \frac{37}{864} - \frac{25}{288} \zeta(3) \right] \Cc_2 \Cc_4
\right. \nonumber \\
&& \left. ~~~~
+ \left[ \frac{1}{16} - \frac{1}{9} \zeta(3) \right] \Cc_{444} 
\right] a^3 ~+~ O(a^4) ~.
\label{genopdim}
\end{eqnarray}
The two loop part is in exact agreement with \cite{32}. Moreover, our two loop
computation was carried out in an arbitrary linear covariant gauge and we
observed the cancellation of the gauge parameter which provides an additional
check. The three loop diagrams were computed in the Feynman gauge and the 
double and triple poles in $\epsilon$ in the three loop renormalization 
constant satisfy the underlying renormalization group formalism as otherwise a 
finite expression would not have emerged. Therefore we are confident that  
(\ref{genopdim}) correctly extends the result of \cite{32}. If one restricted 
to four dimensions then the final term involving $\Cc_{444}$ would be absent 
and the corresponding four dimensional expressions for the remaining $\Cc_i$ 
used.

{\begin{table}[ht]
\begin{center}
\begin{tabular}{|c|c||r|r|r|r|}
\hline
Spin & Chirality & $\Cc_0$ & $\Cc_2$ & $\Cc_4$ & $\Cc_{444}$ \\ 
\hline
$(\half,0)$ & $+$ & 1 & $12$ & $72$ & 0 \\ 
$(\half,0)$ & $-$ & 1 & $12$ & $-$ $24$ & 0 \\ 
$(\threehalves,0)$ & $+$ & 1 & $-$ $12$ & $72$ & 0 \\ 
$(1,\half)$ & $-$ & 1 & $-$ $4$ & $-$ $24$ & 0 \\ 
\hline
\end{tabular}
\end{center}
%\vspace{0.3cm}
\begin{center}
{Table $1$. Values for the evaluation of the general anomalous dimension for 
various nucleons.}
\end{center}
\end{table}}

Equipped with the general anomalous dimension we can extend the two loop 
results in the renormalization scheme of \cite{32} for eigen-operators with
specific spins and chirality. As indicated in \cite{32} these relate to various 
operators in the literature. If we denote the label which the irreducible 
representations of the Lorentz group with two spins $j$ and $\bar{j}$ by
$(j,\bar{j})$ have then the eigen-operators are ${\cal O}^{(j,\bar{j})}$. They
can be written in the forms, \cite{23,32,38,39},
\begin{eqnarray}
{\cal O}_+^{(\half,0)} &=& \epsilon^{IJK} \psi_L^I \left( \left( \psi_L^J
\right)^T C \psi_L^K \right) ~~~,~~~ 
{\cal O}_-^{(\half,0)} ~=~ \epsilon^{IJK} \psi_R^I \left( \left( \psi_L^J
\right)^T C \psi_L^K \right) \nonumber \\
{\cal O}_+^{(\threehalves,0)} &=& \epsilon^{IJK} \Deltaslash \psi_L^I 
\Deltaslash \psi_L^J \Deltaslash \psi_L^K ~~~,~~~ 
{\cal O}_+^{(1,\half)} ~=~ \epsilon^{IJK} \Deltaslash \psi_L^I 
\Deltaslash \psi_L^J \Deltaslash \psi_R^K ~. 
\end{eqnarray}
where $\Delta^2$~$=$~$0$. Here we have denoted right and left handed quarks by 
$\psi_R$~$=$~$\half(1+\gamma^5)\psi$ and $\psi_L$~$=$~$\half(1-\gamma^5)\psi$ 
respectively. To determine the anomalous dimensions of each operator from the
general anomalous dimension we replace these four dimensional tensor product
matrices $\Cc_i$ by their eigenvalue under the Lorentz symmetry, \cite{32}.
These have been given in \cite{32} but are summarized in Table $1$. Hence we 
have 
\begin{eqnarray}
\gamma^{(\half,0)}_+(a) &=& -~ 2 a + \left[ \frac{70}{3} - \frac{16}{9} \Nf 
\right] a^2
+ \left[ \frac{56}{27} \Nf^2 + \left[ \frac{80}{3} \zeta(3) - 80 \right] \Nf 
+ \frac{5392}{9} + 16 \zeta(3) \right] a^3 \nonumber \\
&& +~ O(a^4) \nonumber \\
\gamma^{(\half,0)}_-(a) &=& -~ 2 a + \left[ \frac{50}{3} - \frac{16}{9} \Nf 
\right] a^2
+ \left[ \frac{56}{27} \Nf^2 + \left[ \frac{80}{3} \zeta(3) - \frac{2132}{27} 
\right] \Nf + \frac{5494}{9} \right] a^3 + O(a^4) \nonumber \\
\gamma^{(\threehalves,0)}_+(a) &=& 2 a + \left[ \frac{164}{3} 
- \frac{20}{9} \Nf \right] a^2
+ \left[ \frac{4}{27} \Nf^2
- \left[ \frac{1004}{9} + \frac{80}{3} \zeta(3) \right] \Nf + 1191 
- \frac{560}{3} \zeta(3) \right] a^3 \nonumber \\
&& +~ O(a^4) \nonumber \\
\gamma^{(1,\half)}_-(a) &=& \frac{2}{3} a + \left[ \frac{118}{3} 
- \frac{56}{27} \Nf \right] a^2
+ \left[ \frac{64}{81} \Nf^2
- \left[ \frac{8084}{81} + \frac{80}{9} \zeta(3) \right] \Nf + \frac{9248}{9} 
- \frac{272}{3} \zeta(3) \right] \! a^3 \nonumber \\
&& +~ O(a^4) ~.
\label{ksanomdim}
\end{eqnarray} 
To assist with a comparison the numerical values are
\begin{eqnarray}
\gamma^{(\half,0)}_+(a) &=& -~ 2.0000000 a + \left[ 23.3333333 - 1.7777778 \Nf
\right] a^2 \nonumber \\
&& + \left[ 2.0740741 \Nf^2 - 47.9451492 \Nf + 618.3440216 \right] a^3 ~+~
O(a^4) \nonumber \\
\gamma^{(\half,0)}_-(a) &=& -~ 2.0000000 a + \left[ 16.6666667 - 1.7777778 \Nf
\right] a^2 \nonumber \\
&& + \left[ 2.0740741 \Nf^2 - 46.9081122 \Nf + 610.4444444 \right] a^3 ~+~
O(a^4) \nonumber \\
\gamma^{(\threehalves,0)}_+(a) &=& 2.0000000 a + \left[ 54.6666667 - 2.2222222
\Nf \right] a^2 \nonumber \\
&& + \left[ 0.1481481 \Nf^2 - 143.6104063 \Nf + 966.6160447 \right] a^3 ~+~
O(a^4) \nonumber \\
\gamma^{(1,\half)}_-(a) &=& 0.6666667 a + \left[ 39.3333333 - 2.0740741
\Nf \right] a^2 \nonumber \\
&& + \left[ 0.7901235 \Nf^2 - 110.4874194 \Nf + 918.5690630 \right] a^3 ~+~
O(a^4) ~.
\end{eqnarray}
Clearly the coefficients of the anomalous dimension derived from the
generalized operator approach of \cite{32} are larger in value than our 
$\MSbar$ direct calculation. Therefore it would appear that the latter 
anomalous dimensions have a slower rate of convergence.

Comparing the expressions for the $(\half,0)$ pair of operators to our $\MSbar$
expressions we see that the one loop terms are the same. This is expected since
that part of an anomalous dimension is scheme independent. The two and three
loop terms are not the same. This difference is due to renormalization scheme
dependence. While the subtraction method used is in principle the same in both
cases since only the poles in $\epsilon$ are removed into renormalization
constants, it is in the derivation of these poles in the $d$-dimensional
calculations where the differences arise. Moreover, in \cite{32} the issue of
having to handle $\gamma^5$ in dimensional regularization is circumvented 
unlike our extension of the two loop $\MSbar$ computation of \cite{20} where we
completely reproduced that result. One issue relating to this concerns whether 
it is possible to derive (\ref{anomdim}) from the general operator anomalous 
dimension, (\ref{genopdim}). A clue resides in the comparison of the expression
for both chiralities of the spin $\half$ operators. If we compute the ratio of 
the anomalous dimensions of the chiralities for both spin $\half$ cases we find
\begin{eqnarray}
\frac{\gamma^{(\half,0)}_+(a)}{\gamma_+(a)} &=& 1 ~+~ 
\frac{7}{18} [2\Nf - 33] a ~-~ \frac{1}{162} \left[ 52 \Nf^2 - 4194 \Nf + 34821
\right] a^2 ~+~ O(a^3) \nonumber \\ 
\frac{\gamma^{(\half,0)}_-(a)}{\gamma_+(a)} &=& 1 ~+~ 
\frac{7}{18} [2\Nf - 33] a ~-~ \frac{1}{162} \left[ 52 \Nf^2 - 3774 \Nf + 27891
\right] a^2 ~+~ O(a^3) \,.
\end{eqnarray}
As in \cite{32} the first two terms are in agreement but differ now in the 
$O(a^2)$ terms. Moreover, the coefficient of the one loop $\beta$-function 
appears in the discrepancy between the $\MSbar$ result of (\ref{anomdim}) and 
(\ref{ksanomdim}). This can be explained by recalling that in the mapping of 
the general result $\Cc_2$ is replaced by $12$. However, this is the four 
dimensional evaluation of $d(d-1)$ which is derived from the product 
$\Gamma_{(2)}^{\mu\nu} \Gamma_{(2)\,\mu\nu}$. Expressing this in terms of 
$\epsilon$ gives
\begin{equation}
d (d-1) ~=~ 12 \left[ 1 ~-~ \frac{7}{6} \epsilon ~+~ \frac{1}{3} \epsilon^2
\right] ~. 
\end{equation}
Such a factor emerges from $\Cc_2$ when one projects (\ref{genopdim}) onto the
proton operator by using the formal contraction $I \otimes I$. Therefore, 
including the contribution from the $O(\epsilon)$ term within the mapping used
in \cite{32} reproduces the corresponding two loop terms of (\ref{anomdim}). To
extend this to the next order is certainly highly non-trivial. This is partly 
because there are more terms corresponding to additional operators in 
(\ref{genopdim}) but also due to the now hidden general evanescent operators as
well as the $\gamma^5$ issue. The operators are hidden in the sense that their 
effect in $d$-dimensions cannot be restored from the explicit expression in 
(\ref{genopdim}). More crucially, though, in order to proceed along these lines
one is in effect repeating the actual calculation anyway which was used to 
derive (\ref{anomdim}) in the first place. Indeed an analogous analysis for
$4$-fermi operators, \cite{52}, only serves to illustrate the large complexity
of such a problem which is beyond the scope of the present article.
 
\sect{Discussion.}

By way of concluding remarks we note that first we have extended the two loop
$\MSbar$ renormalization of the $3$-quark proton operator, \cite{20}, to three 
loops. This is a technically more involved computation than \cite{20} since the
$\gamma^5$ problem in dimensional regularization can no longer be treated 
passively at three loops. To accommodate this we have extended Larin's method
for automatic symbolic manipulation programmes to operators which mix under
renormalization. Similar features to \cite{31} emerge in that the finite 
renormalization constant which is required to restore anti-commutativity in
four dimensions is independent of the gauge parameter. We have indicated that
this is true to all orders if the naive anomalous dimensions of the operators
are independent of the gauge parameter which is the case for $\MSbar$. By 
contrast in other mass dependent renormalization schemes this finite 
renormalization would be gauge dependent. The reasoning for this is that when a
gauge invariant operator is renormalized in a mass dependent scheme its 
anomalous dimension depends on the gauge parameter. In addition what is 
apparent from comparing the various finite renormalization constants used to
restore chirality in four dimensions both here and in \cite{31} is that there 
is no {\em universal} finite renormalization. In other words one cannot merely 
extract a $Z^5$ from \cite{31} and use it within another computation where the 
seed operator is not even present. While it may appear to be satisfactory at a 
particular low loop order for a Green's function it will lead to 
inconsistencies at higher loop order. In other words for each appearance of 
$\gamma^5$ within an automatic symbolic manipulation computation one has to 
systematically treat $\gamma^5$ in an adaptation of the Larin method.

Moving away from the above general remarks concerning $\gamma^5$ we note that
the $3$-quark operator we concentrated on was that relating to the proton or
$(\half,0)$ in Lorentz spin notation. Ideally other spin operators are of 
interest and we have provided a first step in that direction by extending the
recent analysis of \cite{32} to three loops. This used a generalized operator
approach which resolved the evanescent and $\gamma^5$ issues from another
angle. Though as we have remarked it is clear that the results are not in the
$\MSbar$ scheme if one regards the earlier work of \cite{31} as the true 
$\MSbar$ situation which we are in agreement with. To extend the $3$-quark
operator $\MSbar$ renormalization to other spin operators is not 
straightforward. This is because there is mixing between operators which is
dependent on flavour symmetry. Our computational setup was designed purely for
the proton case and will need to be extended to accommodate these other 
operators which is a topic we hope to return to later. The explicit definition 
of these operators and the relation to the flavour structure is given in 
\cite{24,39}. Whilst the motivation for this work is in relation to providing 
the perturbative structure of Green's functions to assist lattice matching in 
the high energy limit, in order to refine the understanding of proton structure
will require an extension of our analysis in another direction. In essence this 
involves the treatment of operators with higher moments which manifest 
themselves in the decoration of (\ref{origop}) with covariant derivatives. For 
instance, a three loop renormalization of the first moment will require the 
order of an additional $700$ Feynman diagrams to be calculated. Again we hope 
to return to this in a later analysis.

\vspace{1cm}
\noindent
{\bf Acknowledgement.} The author thanks Dr. R. Horsley and Dr. P.E.L. Rakow
for useful discussions.

\end{document}